\documentclass[aps,pre,reprint,onecolumn,showpacs,groupedaddress]{revtex4-1}
\usepackage{graphicx}% Include figure files
\usepackage{amssymb}
\usepackage{amsmath}
\usepackage{color}
\usepackage{hyperref}

\def\kon{k_{\rm on}}
\def\konc{k_{\rm on}c}
\def\koff{k_{\rm off}}
\def\Kon{K_{\rm on}}
\def\Konc{K_{\rm on}c}
\def\Koff{K_{\rm off}}
\def\mum{$\mu$m}
\def\pers{ s$^{-1}$}

\def\pernm{ nM$^{-1}$}

\begin{document}

\title{Phase-plane analysis of the totally asymmetric simple exclusion
  process with binding kinetics and switching between antiparallel
  lanes}

\author{Hui-Shun Kuan} \affiliation{Department of Chemistry and
  Biochemistry, University of Colorado, Boulder} \author{Meredith
  D. Betterton} \affiliation{Department of Physics, University of
  Colorado, Boulder} \date{\today}

\begin{abstract}
  Motor protein motion on biopolymers can be described by models
  related to the totally asymmetric simple exclusion process
  (TASEP). Inspired by experiments on the motion of kinesin-4 motors
  on antiparallel microtubule overlaps, we analyze a model
  incorporating the TASEP on two antiparallel lanes with binding
  kinetics and lane switching. We determine the steady-state motor
  density profiles using phase plane analysis of the steady-state mean
  field equations and kinetic Monte Carlo simulations. We focus on the
  the density-density phase plane, where we find an analytic solution
  to the mean-field model. By studying the phase space flows, we
  determine the model's fixed points and their changes with
  parameters. Phases previously identified for the single-lane model
  occur for low switching rate between lanes. We predict a new
  multiple coexistence phase due to additional fixed points that
  appear as the switching rate increases: switching moves motors from
  the higher-density to the lower-density lane, causing local jamming
  and creating multiple domain walls.  We determine the phase diagram
  of the model for both symmetric and general boundary conditions.
\end{abstract}

\maketitle

\section{Introduction}

Motor protein motion along biological polymers is important for many
biological processes \cite{bray_cell_2000}. Examples include kinesin
walking along microtubules and ribosomes moving along mRNA
\cite{kolomeisky_motor_2015,chowdhury_modeling_2013}. These filaments
act as one-dimensional lanes that allow proteins to move over long
distances and accumulate at the correct location for their biological
function. Physical models of motor protein motion often incorporate
two main features: directional motion along a filament and
binding/unbinding.

The directional motion of motor proteins is a remarkable
implementation of a classic model of diven-diffusive transport, the
totally asymmetric simple exclusion process (TASEP)
\cite{_nonequilibrium, helbing_traffic_2001}. In the TASEP, particles
move unidirectionally along a one-dimensional lattice and experience
excluded volume interactions. The TASEP and its variants have been
applied to one-dimensional nonequilibrium transport problems ranging
from molecular motors to vehicular and pedestrian traffic. In contrast
to thermodynamic systems, the non-equilibrium steady-state solution of
the TASEP is sensitive to the boundary conditions, even in the bulk of
the lane \cite{krug_boundaryinduced_1991, _nonequilibrium,
  _statistical_1995-1, chou_nonequilibrium_2011}. The TASEP has been
solved exactly by Derrida et al.~\cite{derrida_exact_1993}. Three
phases can occur, the low-density, high-density, and maximum current
states. Kolomeisky et al.~\cite{kolomeisky_phase_1998} analyzed the
formation of the steady-state phases in the mean-field equation by
analyzing the dynamics of domain walls that can appear when two phases
coexist in the same lane. This work also found that the boundary
conditions are not always satisfied and there is no steady localized
domain wall in the pure TASEP \footnote{More precisely, there is a
  possibility for a localized domain wall to appear when the inward
  current and outward currents are equal. However, the domain wall can
  appear at any point. Thus, the average density profile becomes a
  line instead of the high density-low density coexistence phase.}.

Because binding kinetics are important for most motor proteins,
biohysical models have extended the TASEP to include motor binding and
unbinding (Langmuir kinetics, LK). Parmeggiani, Franosch, and Frey
(PFF) developed a single-lane TASEP plus LK model and determined the
mean-field solutions
\cite{parmeggiani_phase_2003,parmeggiani_totally_2004}. They
discovered a new phase with low density-high density coexistence in
this model, implying that domain wall localization can occur due to
LK.  Experimental work measured kinesin-8 motor protein traffic jams
on stabilized microtubules, and found good agreement with the density
profiles predicted by PFF \cite{leduc_molecular_2012}.

TASEP-inspired models have been applied to motor proteins that move on
cytoskeletal filaments and affect filament length. Motors can affect
the lengths of microtubules \cite{varga_kinesin8_2009}, antiparallel
microtubule overlaps \cite{bieling_minimal_2010}, and the
microtubule-based the mitotic spindle
\cite{goshima_length_2005,walczak_xkcm1_1996}. Kinesin-8 motors walk
with directional bias and and promote microtubule plus-end shortening
\cite{gupta_endspecific_2006,varga_yeast_2006,varga_kinesin8_2009}. These
experiments have inspired theory to describe how length-dependent
depolymerization affects otherwise static microtubles
\cite{hough_microtubule_2009,varga_kinesin8_2009,reese_crowding_2011},
microtubules with simplified polymerization kinetics,
\cite{govindan_length_2008,johann_length_2012,melbinger_microtubule_2012,reese_molecular_2014},
and dynamic microtubules
\cite{tischer_providing_2010,kuan_biophysics_2013,gluncic_kinesin8_2015}.

The bipolar structure of the mitotic spindle leads to overlapping
antiparallel microtubules at the center of the spindle.  Control of
microtubule overlaps is therefore important for mitosis and
cytokinesis.  Microtubule (MT) crosslinking (by PRC1/Ase1/MAP65) and
motion of kinesin-4 motors (chromokinesins) stabilize MT
antiparallel overlaps
\cite{kurasawa_essential_2004,zhu_cell_2005,khmelinskii_cdc14regulated_2007},
along with other motors and proteins
\cite{fededa_molecular_2012,subramanian_marking_2013}. Bieling,
Telley, and Surrey (BTS) reconstituted a minimal system of stable
antiparallel MT overlaps in which the crosslinking protein PRC1 bound
preferentially to overlapping regions of antiparallel MTs
\cite{bieling_minimal_2010}. PRC1 recruited the kinesin-4 motor Xklp1
to the overlap. Xklp1 motors could bind to and unbind from the MTs,
walk toward the plus end of each MT, and switch between the two MTs at
a relatively high rate \cite{bieling_minimal_2010}. Motors present
near the MT plus ends slowed the polymerization speed, consistent with
earlier work showing that Xklp1 inhibits dynamic instability
\cite{bringmann_kinesinlike_2004} and affects spindle MT mass
\cite{castoldi_chromokinesin_2006}. As a result, antiparallel MT
overlaps reached a constant length that depended on the bulk
concentration of motors.  This work demonstrated that motor-dependent
regulation of dynamics and length can occur not just for single MTs,
but for overlapping MT pairs.

Recently we developed a model inspired by the BTS experiments in which
we studied antiparallel lanes with TASEP, LK, and lane switching for
fixed-length lanes \cite{kuan_motor_2015}. Our work is related to
previous generalizations of the TASEP to multiple lanes and coupling
between lanes.  Multi-lane systems with two or more species have been
studied \cite{pronina_twochannel_2004, xiao_asymmetric_2009,
  gupta_asymmetric_2014, levine_spontaneous_2004,
  pronina_twochannel_2004, chai_traffic_2009, ashwin_queueing_2010}.
Reichenbach et al. \cite{reichenbach_exclusion_2006} studied parallel
and Juh\'asz \cite{juhasz_weakly_2007} anti-parallel lanes without LK;
both derived analytic solutions. In these models, even though LK is
absent, domain wall localization can occur due to switching events
which balance the flux (in the work of PFF, this is called the
matching condition \cite{parmeggiani_totally_2004}).  Multi-lane
models that included LK were studied by Gupta and Dhiman
\cite{gupta_asymmetric_2014} (parallel lanes), and Levine and Willmann
\cite{levine_spontaneous_2004} (antiparallel lanes). This work found
that multiple phases appear and that analytic solutions can be derived
in some limites. Other related work includes that of Chai et
al. \cite{chai_traffic_2009}, who studied multiple species on one lane
with some non-moving species, and Nowak et
al. \cite{nowak_dynamic_2007}, who studied fluctuating boundary
conditions.  Pierobon et al. \cite{pierobon_bottleneckinduced_2006}
considered the case in which the lane has a defect, which makes a
singular point in the density profile, causing new bottleneck phases
to appear.

In our previous paper, we compared steady-state density profiles of
our model to those determined experimentally, and discussed how the
appearance of a localized domain wall can be understood a using total
binding constraint \cite{kuan_motor_2015}.  Here we extend our
previous work by analyzing the density-density phase plane to solve
the steady-state mean-field equations and determine the phase
boundaries. Analyzing the model's phase-space flows and fixed points,
as well as their changes with parameters, allows us to calculate the
phase diagram.  Some previous work has discussed fixed-points of TASEP
models \cite{parmeggiani_totally_2004,levine_spontaneous_2004}. Yadav
et al. used phase-plane analysis of a fixed-point-based boundary layer
method to study multi-lane TASEP models \cite{yadav_phase_2012}. Here
we undertake a detailed study of the model's phase-space flows and
fixed points and along with an analytic phase-plane solution. This
allows us both to develop intuition and calculate the mean-field phase
diagram with minimal assumptions.  We explain why high motor switching
rate between the two lanes leads to a new low density-high density-low
density-high density coexistence phase, for which multiple domain
walls occur in the bulk of the system.  We also extend our previous
work, which only considered symmetric boundary conditions
\cite{kuan_motor_2015}, to the case of asymmetric boundary conditions.

In sec.~\ref{sec:model}, we describe the discrete model and derive the
mean-field approximation.  Using the random phase approximation and
Taylor expansion, we derive the steady-state mean-field differential
equations. Then in sec.~\ref{sec:phase}, we develop a method to derive
key features of the density profiles using phase-space flows. This is
a different approach from determining the position-dependent density
profiles that were the focus of previous work
\cite{parmeggiani_totally_2004, levine_spontaneous_2004,
  reichenbach_exclusion_2006, xiao_asymmetric_2009,
  gupta_asymmetric_2014, levine_spontaneous_2004,
  pronina_twochannel_2004, chai_traffic_2009}.  In
sec.~\ref{sec:sym_phase_diagram}, we determine the nonlinear phases of
the model with symmetric boundary conditions. We also derive an
analytic approximation to the position-dependent solution and two ways
to determine domain wall positions and phase boundaries. In
sec.~\ref{sec:phase_diagram}, we determine the phase diagram for
symmetric boundary conditions, and in sec.~\ref{sec:general_case}
discuss the general case of asymmetric boundary
conditions. Sec.~\ref{sec:conclusion} is the conclusion.

\section{Model}
\label{sec:model}

\begin{figure}[t!]
\begin{centering}
\includegraphics[width=0.45 \textwidth]{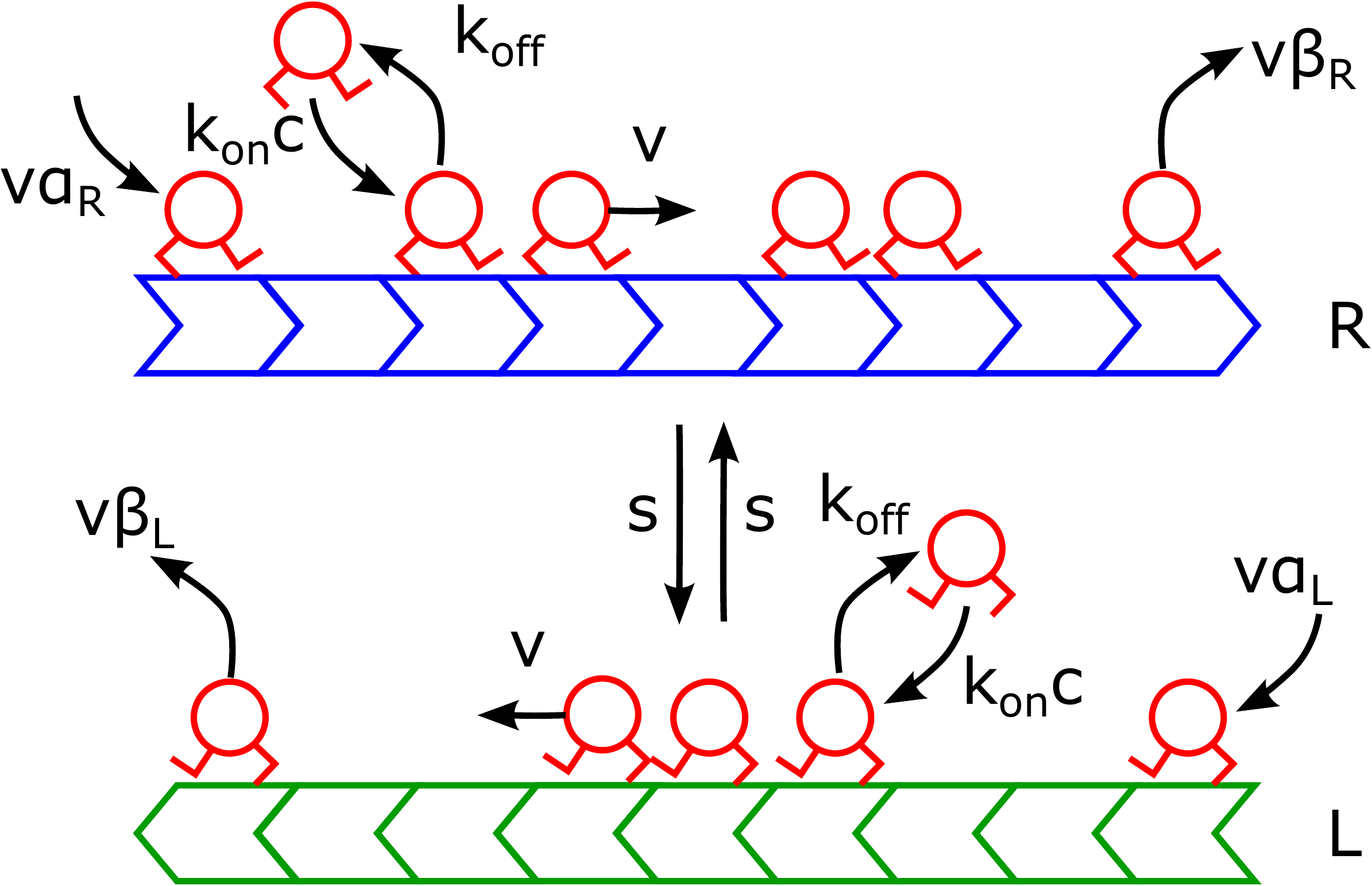}
\caption{Schematic of the antiparallel two-lane TASEP with Langmuir
  kinetics and lane switching.  Two lanes (green and blue) have their
  plus ends (indicating the direction of motor motion) oppositely
  oriented. The blue lane (R) has plus end to the right; the green
  lane (L) has plus end to the left. Motors (red) bind to empty
  lattice sites with rate $\konc$, where $c$ is the bulk motor
  concentration, and unbind with rate $\koff$. Bound motors step
  toward the lane plus end with rate $v$ (if the adjacent site toward
  the plus end is empty) or switch to the other lane with rate $s$
  (if the corresponding site on the adjacent lane is empty). At
  minus ends, motors are inserted at rate $\alpha_{R, L} v$.  At plus
  ends, motors are removed at rate $\beta_{R, L} v$.}
\label{cartoon}
\end{centering}
\end{figure}

Our model of motor motion on antiparallel lanes \cite{kuan_motor_2015}
is based on the BTS experiments \cite{bieling_minimal_2010}.  Motors
move toward lane plus ends, bind to and unbind from each lane, and
switch between lanes (fig.~\ref{cartoon}).  We study lanes with fixed
number of sites $N$. At each site, motor binding occurs with binding
rate $\konc$, where $\kon$ is the binding rate constant per site and
$c$ the bulk motor concentration, and motor unbinding occurs with rate
$\koff$. Each bound motor steps at rate $v$ to the next site toward
the lane plus end (if the next site is unoccupied), and switches at
rate $s$ to the site on the adjacent lane (if that site is
unoccupied).  Nontrivial competition between the motor stepping
(TASEP) and Langmuir kinetics occurs when the overall binding rate to
one lane $\Konc = N\konc$ and unbinding rate $\Koff = N\koff$ are of
similar magnitude to the motor speed $v$
\cite{parmeggiani_totally_2004}.

The discrete model is based on the occupation number $\hat{n}_i$,
which is $1$ (0) if site $i$ is occupied (empty).  For bulk sites
($2<i<N-1$) on lanes with plus end right (R) and left (L) and a small
time increment $\Delta t$, the equations are
\begin{eqnarray}
  \nonumber
  \frac{\hat{n}_{R, i}(t+\Delta t)-\hat{n}_{R, i}(t)}{\Delta t} =&& v\hat{n}_{R, i-1}(t)[1-\hat{n}_{R,
                                     i}(t)]-v\hat{n}_{R,
                                     i}(t)[1-\hat{n}_{R, i+1}(t)] +\konc
                                     [1-\hat{n}_{R, i}(t)] \\ 
                                  &&-\koff \hat{n}_{R, i}(t) - s \hat{n}_{R,
                                     i}(t)[1-\hat{n}_{L, i}(t)] + s
                                     \hat{n}_{L, i}(t)[1-\hat{n}_{R, 
                                     i}(t)],\\ 
  \nonumber
  \frac{\hat{n}_{L, i}(t+\Delta t)-\hat{n}_{L, i}(t)}{\Delta t} =&& v\hat{n}_{L, i+1}(t)[1-\hat{n}_{L,
                                     i}(t)]-v\hat{n}_{L,
                                     i}(t)[1-\hat{n}_{L, i-1}(t)] +\konc
                                     [1-\hat{n}_{L, i}(t)] \\ 
                                  &&-\koff \hat{n}_{L, i}(t) - s \hat{n}_{L,
                                     i}(t)[1-\hat{n}_{R, i}(t)] +
                                     s\hat{n}_{R, i}(t)[1-\hat{n}_{L, 
                                     i}(t)].
\label{Fock_space_formalism}
\end{eqnarray}

The boundary site equations include fluxes into and out of the lanes.
The entering flux is $v \alpha_{R,L}[1-\hat{n}_1(t)]$, and the exiting
flux is $v \beta_{R,L} \hat{n}_N(t)$. We neglect binding and switching
kinetics at the boundary sites. Then we have
\begin{eqnarray}
  \frac{\hat{n}_{R, 1}(t+\Delta t)-\hat{n}_{R, 1}(t)}{\Delta t}&=& v\alpha_R[1-\hat{n}_{R, 1}(t)]-
     v\hat{n}_{R, 1}(t)[1-\hat{n}_{R, 2}(t)],\\ 
  \frac{\hat{n}_{R, N}(t+\Delta t)-\hat{n}_{R, N}(t)}{\Delta t} &=& v\hat{n}_{R,
                                     N-1}(t)[1-\hat{n}_{R,
                                     N}(t)]-v\beta_R\hat{n}_{R, 
     N}(t),\\ 
  \frac{\hat{n}_{L, 1}(t+\Delta t)-\hat{n}_{L, 1}(t)}{\Delta t} &=& v\hat{n}_{L, 2}(t)[1-\hat{n}_{L,
                                     1}(t)]-v\beta_L\hat{n}_{L, 
     1}(t), \\
  \frac{\hat{n}_{L, N}(t+\Delta t)-\hat{n}_{L, N}(t)}{\Delta t} &=& v\alpha_L[1-\hat{n}_{L, N}(t)]-
     v\hat{n}_{L, N}(t)[1-\hat{n}_{L, N-1}(t)].
\label{Fock_space_formalism_boundaries}
\end{eqnarray}
These boundary conditions fix the motor densities to be $\alpha_{R,L}$
at the minus end and $1-\beta_{R,L}$ at the plus end of each lane.

We performed kinetic Monte Carlo (kMC) simulations of the discrete
model with time step $\Delta t$. We applied the following rules for an
overlap with $N$ sites per lane.
\begin{enumerate}
\item Randomly choose a lane (R/L) and site $i$.
\item If the site is empty, attach a motor with probability
  $\konc \Delta t$.  If the site is occupied, detach the motor with
  probability $\koff \Delta t$.
\item If the site is occupied and the adjacent site toward the plus
  end is empty, move the motor forward with probability $v \Delta t$.
\item If the site is occupied and the corresponding site on the
  neighboring lane is empty, switch the motor to the other lane with
  probability $s \Delta t$.
\item Enforce the boundary conditions: site $1$ on lane R and site $N$
  on lane L are occupied with probability $\alpha_{R,L} $, while site
  $N$ on lane R and site 1 on lane L are occupied with probability
  $(1-\beta_{R,L}) $.
\item Repeat steps 1-5 $2N$ times total to sample all sites on both
  lanes.
\end{enumerate}
We chose $\Delta t$ to give a characteristic time for motor
binding/unbinding of about $10^5$ time steps. For the reference
parameter set and a bulk motor concentration of 200 nM, we used
$\Delta t = 5 \times 10^{-4}$. Approximately $4 \times 10^7$ time
steps were used to reach steady state, then $2 \times 10^6$ time steps
were used to collect data. To determine the average motor
concentration, we averaged $10^4$ samples separated by 200 time
steps. The reference parameter set was obtained from the BTS
measurements \cite{bieling_minimal_2010} or estimated based on
comparison of our kMC simulations to the BTS data
\cite{kuan_motor_2015} (table \ref{tab:units}).

\begin{table}%[htbp]
\footnotesize
  \begin{center}
    \begin{tabular}{|cp{4cm}p{3.5cm}p{6cm}|}
      \hline
      Symbol & Parameter & Reference value &  Notes  \\
      \hline
      $v$  & Motor speed & 0.5 $\mu$m\pers &   Measured by
                                             \citet{bieling_minimal_2010}\\  
      $\kon$  & Binding rate constant   &
                                          $2.7\times10^{-4}$
                                          \pernm\pers &   Estimated
                                                        based on motor
                                                        density
                                                        profiles and
                                                        kymographs in
                                                        \citet{kuan_motor_2015} \\ 
      $c$  & Bulk motor concentration & 1--200 nM &  Varied
                                                    by
                                                    \citet{bieling_minimal_2010}\\ 
      $\koff$  & Unbinding rate &0.169\pers &  Measured
                                              by
                                              \citet{bieling_minimal_2010} \\
      $s$  & Switching rate & $0.44$ \pers& Measured by \citet{bieling_minimal_2010} \\  
      $\alpha$  &Motor flux constant into overlap from MT minus end & 0 &
                                                                          Motors
                                                                          bind
                                                                          primarily
                                                                          inside
                                                                          the
                                                                          overlap;
                                                                          see
                                                                          discussion
                                                                          in
                                                                          \citet{kuan_motor_2015}. We
                                                                          varied
                                                                          $\alpha$
                                                                          between
                                                                          0 and 1 to determine the model phase diagram\\  
      $\beta$  &Motor flux constant out of overlap from MT plus end & 0 &
                                                                          An
                                                                          upper
                                                                          bound
                                                                          on
                                                                          the
                                                                          end
                                                                          motor
                                                                          unbinding
                                                                          rate is $\beta =
                                                                          2.7 \times
                                                                          10^{-3}$;
                                                                          see
                                                                          \citet{kuan_motor_2015}. We varied
                                                                          $\beta$
                                                                          between
                                                                          0 and 1 to determine the model phase diagram\\   
      $N$    & Number of sites & 1000& We
                                       used $N=1000$ unless
                                       otherwise specified \\
      \hline
    \end{tabular}
    \caption{Parameter values for the reference parameter set, taken
      from experimental measurements or estimated as noted.}
    \label{tab:units}
  \end{center}
\end{table}

\subsection{Mean-field continuum model}
\label{sec:meanfield}

We derived the mean-field continuum approximation to the model as in
previous work \cite{parmeggiani_totally_2004, kuan_motor_2015}. We
applied the stationary average
$\langle\hat{n}_i\rangle \equiv \rho_i$, the random phase
approximation
$\langle\hat{n}_i\hat{n}_{i+1}\rangle =
\langle\hat{n}_i\rangle\langle\hat{n}_{i+1}\rangle$,
assumed motor commutation during switching
$\langle \hat{n}_{R,i} \hat{n}_{L,i}\rangle =\langle
\hat{n}_{L,i}\rangle \langle \hat{n}_{R,i}\rangle$,
and time derivative
$\frac{\langle\hat{n_i(t+\Delta
    t)}\rangle-\langle\hat{n_i(t)}\rangle}{\Delta t} \to \frac{d
  \langle\hat{n(t)}_i\rangle}{dt} = \frac{d \rho_i}{dt}$
by taking $\Delta t \rightarrow 0$.  We then Taylor expanded to take
the continuum limit and nondimensionalized by choosing the length of
the overlap, $L$, as the unit of length and $L/v$ as the unit of
time. Capital letters denote the nondimensionalized parameters
($S=sL/v$, etc.). The position variable is changed from site index $i$
to the position variable $x$ and is ranging from $-0.5$ to $0.5$. With
$x=0$ the center of the overlap, the boundary conditions become
$\rho_R(x=-\frac{1}{2})=\alpha_R$, $\rho_L(x=\frac{1}{2}) = \alpha_L$
and $\rho_R(x=\frac{1}{2})=1-\beta_R$,
$\rho_L(x=-\frac{1}{2})=1-\beta_L$. The steady-state continuum
mean-field equations are then
\begin{eqnarray}
\label{eq:cont1}
0&=& (2\rho_R-1) \frac{\partial \rho_R}{\partial x} + \Konc (1-\rho_R)
     - \Koff \rho_R -S \rho_R +S \rho_L , \\
\label{eq:cont2}
0&=&(1-2\rho_L) \frac{\partial \rho_L}{\partial x} + \Konc (1-\rho_L)
     - \Koff \rho_L +S \rho_R -S \rho_L. 
\end{eqnarray}
Because the equations are first-order differential equations, only one
boundary condition is required for each. Since each end of the lane
has two boundary conditions, the equations are overdetermined. The
nonlinearities in these equations have a similar form to those of
Burgers' equation in fluid dynamics. Burgers' equation also becomes
overdetermined in the inviscid limit in which terms with second-order
derivatives are neglected, which leads to the formation of shocks or
domain walls which match solutions satisfying the two different
boundary conditions \cite{whitham1974linear, popkov_localization_2003,
  parmeggiani_totally_2004}.  Here we denote $x_w$ the domain wall
position where the solution that obeys the left boundary condition
matches the solution that obeys the right boundary condition. The
matching condition at the domain wall is continuity in the flux
$j(x) = \rho(x)(1-\rho(x))$, which can be written
$j(x_w-\epsilon)=j(x_w+\epsilon)$, where $\epsilon$ is
infinitesimal. Since at the domain wall the density is not continuous,
fulfilling the matching condition requires a density jump
\cite{parmeggiani_totally_2004} of the form
$\rho(x_w-\epsilon)=1-\rho(x_w+\epsilon)$.

\subsection{Total binding constraint}

These equations satisfy a total binding constraint at steady state
found by summing over all sites on both lanes
\cite{kuan_motor_2015}. In the discrete equations, the flux terms of
the form $\hat{n}_{i-1}(t)[1-\hat{n}_{i}(t)]$ sum to zero and only the
binding and boundary terms remain:
\begin{equation}
  \sum_{i=2}^{N-1} [2\konc-(\konc+\koff)(\hat{n}_{R,i} +
  \hat{n}_{L,i})] +v\alpha_R[1-\hat{n}_{R, 1}(t)]+v\alpha_L[1-\hat{n}_{L,
    N}(t)]-v\beta_R\hat{n}_{R,N}(t)-v\beta_L\hat{n}_{L, 1}(t)=0. 
\label{global_property}
\end{equation}
This gives a constraint on the summed motor occupancy
\begin{equation} 
  \sum_{i=1}^N \hat{n}_{R,i} +\hat{n}_{L,i} = 2 N
  \rho_0 +
  \frac{v[\alpha_R(1-\alpha_R)+ \alpha_L(1-\alpha_L)-
    \beta_R(1-\beta_R)-\beta_L(1-\beta_L)]}{\konc+\koff},   
  \label{eq:constraint}
\end{equation}
where we have defined the Langmuir density
$\rho_0=\konc/(\konc+\koff)$. Therefore, at steady state an effective
binding equilibrium that reflects binding, unbinding, and the
lane-end boundary conditions must be reached on average for the
entire system. This is related to the zero-current condition found in
previous work on the two-lane antiparallel TASEP without binding
kinetics \cite{juhasz_weakly_2007,ashwin_queueing_2010}.

In the continuum mean-field model, the total binding constraint
becomes
\begin{equation}
  \label{eq:contin_constr}
  \int_{-\frac{1}{2}}^{\frac{1}{2}}dx\ \rho_R(x)+ \rho_L(x) =  
    2 \rho_0 + \frac{ \alpha_R(1-\alpha_R) -
    \beta_R(1-\beta_R)+\alpha_L(1-\alpha_L) -
    \beta_L(1-\beta_L)}{\Kon c + \Koff}. 
\end{equation}

\section{Phase plane solution}
\label{sec:phase}

One solution to the steady-state mean field equations \eqref{eq:cont1}
and \eqref{eq:cont2} is the constant solution at the Langmuir density
$\rho_0=\Konc/(\Konc+\Koff)$.  To study spatially varying solutions,
we define the differences of the densities from $\frac{1}{2}$, 
$\sigma_{R,L}(x) = \rho_{R,L}(x) - \frac{1}{2}$. The equations can
then be written
\begin{eqnarray}
\frac{d \sigma_R}{d x} &=& \frac{k}{2} - \frac{\gamma}{4
   \sigma_R}-\frac{S \sigma_L}{2 \sigma_R},
\label{flow_q+}\\ 
\frac{d \sigma_L}{d x} &=& -\frac{k}{2} + \frac{\gamma}{4
   \sigma_L}+\frac{S \sigma_R}{2 \sigma_L}, 
\label{flow_q-}
\end{eqnarray}
where we have defined the rate combinations $k = \Konc+\Koff+S$ and
$\gamma = \Konc-\Koff$, and equations \eqref{flow_q+} and
\eqref{flow_q-} are well defined for $\sigma_{R,L}\neq 0$.  We have
not determined $x$-dependent expressions for $\sigma_R(x)$ and
$\sigma_L(x)$ by solving these equations. Instead, we determined an
implicit solution by first defining the sum and difference of the
densities, $\phi(x)=\sigma_R+\sigma_L$ and
$\omega(x) = \sigma_R-\sigma_L$.  The equations become
\begin{eqnarray}
\label{eq:pi}
  \frac{d \phi}{d x} &=& \frac{\gamma \omega +2S \phi
                         \omega}{\phi^2-\omega^2},\\ 
  \frac{d \omega}{d x} &=& \frac{(k-S)\phi^2-\gamma \phi -
                           (k+S)\omega^2}{\phi^2-\omega^2},   
\label{eq:QQ_delta}
\end{eqnarray}
which combine to give
\begin{equation}
  \label{eq:1}
  \omega \frac{d \omega}{d \phi} = \frac{(k-S)\phi^2-\gamma \phi
    -(k+S)\omega^2}{\gamma +2S \phi }.
\end{equation}
Defining $\eta(\phi)=\omega^2(\phi)$, this can be rewritten
\begin{equation}
  \label{eq:eta}
  \frac{1}{2} \frac{d \eta}{d \phi} = \frac{(k-S)\phi^2-\gamma \phi
    -(k+S)\eta}{\gamma +2S \phi },
\end{equation}
or
\begin{equation}
  \frac{\gamma +2S \phi }{2} d \eta + [\gamma
  \phi-(k-S)\phi^2+(k+S)\eta]d \phi = 0. 
\end{equation}
This inexact ODE can be made exact through multiplication by the
integrating factor $(\gamma+2S\phi)^{k/S}$. We then obtain the
solution by direct integration,
\begin{eqnarray}
  C_1&=&\int d \eta\ (\gamma+2S\phi)^{k/S} \left[\frac{\gamma +2S \phi
         }{2}\right] + \int d \phi\ (\gamma+2S\phi)^{k/S} \left[\gamma 
         \phi-(k-S)\phi^2+(k+S)\eta\right],\\ 
  C_1&=&\frac{\eta}{2} (\gamma+2S\phi)^{1+k/S}  -
         (\gamma+2S\phi)^{1+k/S} \frac{\gamma^2-2(k+S)\gamma\phi
         +(k-S)(k+2S)\phi^2 }{2(k+2S)(k+3S)},
\end{eqnarray}
which gives the solution
\begin{equation}
  \omega^2(\phi)=\frac{C}{(\gamma+2S\phi)^{1+k/S}} -
                    \frac{2(k+S)\gamma\phi -(k-S)(k+2S)\phi^2
                    -\gamma^2}{(k+2S)(k+3S)}.
\label{exact}
\end{equation}
Here $C_1$ and $C$ denote integration constants.  Equation
\eqref{exact} gives solutions for the density profiles in the
$\omega$--$\phi$ or $\sigma_R$--$\sigma_L$ plane.

The integration constant $C$ can be obtained by plugging in the
boundary conditions: $\sigma_R(x=-\frac{1}{2}) = \alpha_R-\frac{1}{2}$,
$\sigma_R(x=\frac{1}{2}) = \frac{1}{2}-\beta_R$, $\sigma_L(x=-\frac{1}{2})=\frac{1}{2}-\beta_L$, and
$\sigma_L(x=\frac{1}{2})=\alpha_L-\frac{1}{2}$.  In much of this paper, we focus on
the symmetric case for which $\alpha_R=\alpha_L=\alpha$ and
$\beta_R=\beta_L=\beta$. Later in section \ref{sec:general_case} we
discuss the general case when $\alpha_R \neq \alpha_L$ and
$\beta_R \neq \beta_L$.

\subsection{Position-dependent approximate solutions}
In equations \eqref{flow_q+} and \eqref{flow_q-}, position-dependent
solutions can be derived by integrating
\begin{eqnarray}
  \nonumber
  \frac{d \sigma_R}{d x} &=& \frac{2 k \sigma_R-\gamma-2 S
                             \sigma_L(x)}{4\sigma_R(x)}\\ 
  \Rightarrow dx &=& \frac{4\sigma_R}{2 k \sigma_R-\gamma-2 S \sigma_L(x)}
                     d\sigma_R. \label{eq:direct}
\end{eqnarray}
Since $\sigma_L$ depends on $x$, this equation is difficult to
integrate directly. However, since equation \eqref{exact} gives the
relationship between $\sigma_R$ and $\sigma_L$, we can rewrite
$\sigma_L(x) = \sigma_L(\sigma_R)$. If we define
$Y_R = \sigma_R-\frac{\gamma}{2k}$, the equation \eqref{eq:direct}
can be written
\begin{eqnarray}
  dx &=& \frac{4}{2
         k}\left(\frac{Y_R+\frac{\gamma}{2k}}{Y_R-\frac{S}{k}
         \sigma_L(Y_R)} \right) d Y_R. 
\label{real_space_q+}
\end{eqnarray}
This allows us to perform direct integration with an appropriate
expansion of $\sigma_L(Y_R)$.

\subsection{Phase space flow and fixed points}
\label{sec:Phase_space_flow_to_determine_the_density_profiles}

\begin{figure}[t!]
\begin{centering}
\includegraphics[width=0.45 \textwidth]{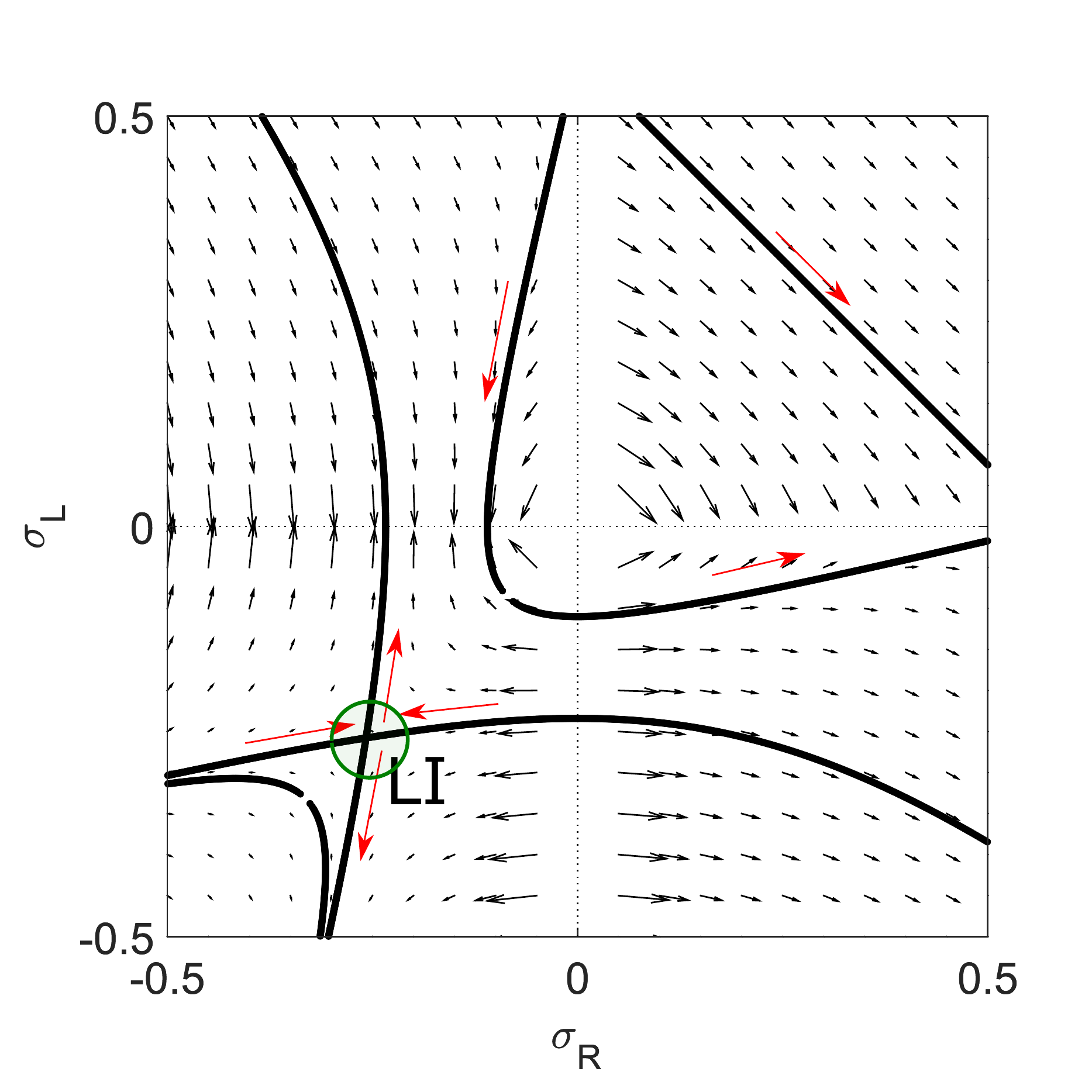}
\includegraphics[width=0.45 \textwidth]{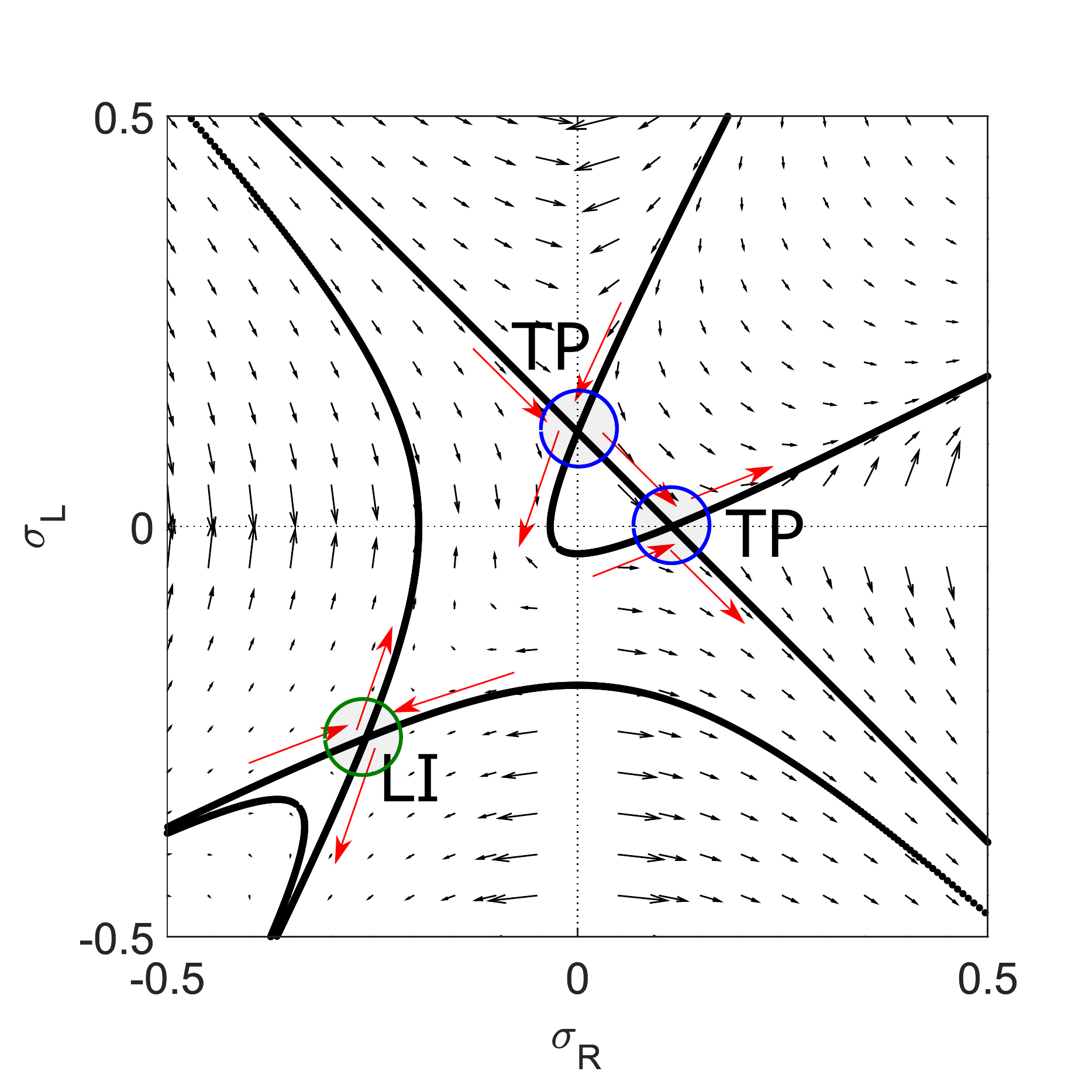}
\caption{Flows in the $\sigma_R$--$\sigma_L$ phase plane for low
  (left, $0.1$\pers) and high (right, $0.5$\pers) switching
  rate. Solid lines are trajectories that pass through the fixed
  points, as discussed in the text.  The Langmuir isotherm is labeled
  LI and the transition points TP. The bulk motor concentration is
  $200$ nM, the motor speed is $5$ \mum\pers, and other parameters are
  the reference values of table \ref{tab:units}. Arrows indicate vector field, which has the mathematical form in \eqref{flow_q+} and \eqref{flow_q-}.}
\label{nonlinear_phase_space_flow}
\end{centering}
\end{figure}

We can determine important features of the density profiles by
studying equations \eqref{flow_q+} and \eqref{flow_q-} in the
$\sigma_R$--$\sigma_L$ phase plane and determining the phase space
flows.  In the phase plane, equations \eqref{flow_q+} and
\eqref{flow_q-} define an effective velocity field that shows the
local change in $\sigma_R$ and $\sigma_L$ at each point in the plane
(fig. \ref{nonlinear_phase_space_flow}). Note that because the
equations are unchanged under the operation $R \to L$, $x \to -x$, the
phase field is symmetric under reflection about the line
$\sigma_L=\sigma_R$.

The flow trajectories are controlled by the fixed points in the phase
plane.  There can be as many as three fixed points: the Langmuir
isotherm (LI), and two transition points (TP) that appear for
sufficiently high switching rate.  Figure
\ref{nonlinear_phase_space_flow} shows the fixed points for low
switching rate (left) and high switching rate (right).  To determine
the fixed points, we rearrange equation \eqref{exact} to solve for the
integration constant:
\begin{equation}
  C = (\gamma+2S\phi)^{1+k/S} \left[\omega^2+\frac{2(k+S)\gamma\phi
      -(k-S)(k+2S)\phi^2 -\gamma^2}{(k+2S)(k+3S)}\right].
\label{exact_2}
\end{equation}

The transition line and points can be determined by the trajectories
with $C = 0$. In this case, either the first term
$(\gamma+2S\phi)^{1+k/S}$ or the second term in square brackets is
zero. If the first term is zero, then
$\gamma+2S(\sigma_R+\sigma_L) = 0$.  This is a line with slope $-1$ in
the $\sigma_R$--$\sigma_L$ plane that intercepts $\sigma_L=0$ at the
point $\sigma_R=(\Koff-\Konc)/(2S)$.  This line can also be derived by
setting equation \eqref{eq:pi} to zero. This is equivalent to
requiring that the total density $\phi=\sigma_R+\sigma_L$ be
independent of $x$; in this case
$\omega(\gamma+2S\phi)]/(\phi^2-\omega^2)=0$, leading to
$\gamma+2S(\sigma_R+\sigma_L) = 0$ as above. Physically, this means
that the switching and binding terms balance. This line is called the
transition line, and the transition points occur where this line
crosses the $\sigma_R = 0$ and $\sigma_L = 0$ lines.  Since the flow
values are ill-defined at the transition points (one of
$d\sigma_{R,L}/dx$ is ill-defined), they can only lie on
$\sigma_{R,L} = 0$ lines.

\begin{figure}[t!]
\begin{centering}
\includegraphics[width=0.45 \textwidth]{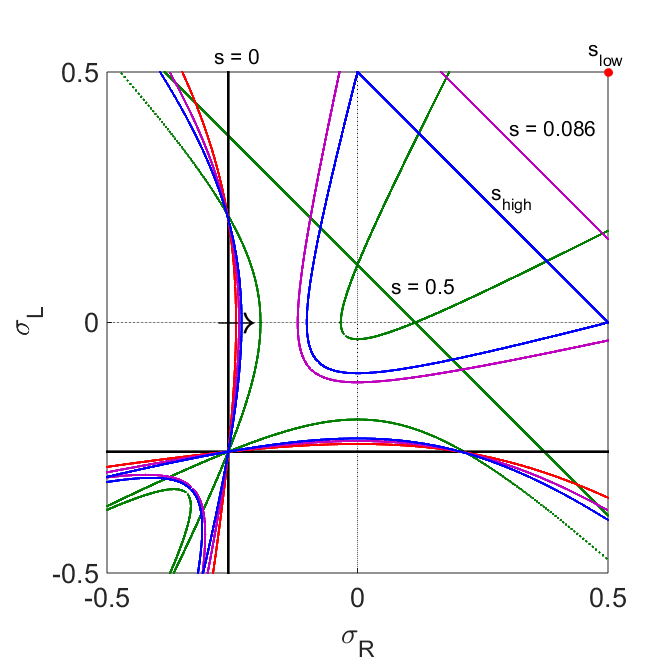}
\caption{Changes in trajectories and fixed points with varying
  switching rate. Curves with $s = 0$\pers, black;
  $s = \frac{3(\koff-\konc)}{4} \approx 0.0862$\pers, purple;
  $s = s_{\rm high} \approx 0.1150$\pers, blue; $s = 0.5$\pers,
  green. The red point labels $s_{\rm low}$.  The arrow indicates how
  the curve with $C_{\rm LI}$ changes as $s$ increases. The bulk motor
  concentration is $c = 200$ nM and motor speed is $5$ \mum\pers;
  other parameters are the reference values of table \ref{tab:units}.}
\label{fixed_points}
\end{centering}
\end{figure}

If the second term in square brackets is zero, the solution is a
hyperbola that satsifies
\begin{equation}
  \omega^2-\frac{k-S}{k+3S}\left[\phi-\frac{\gamma(k+S)}{(k-S)(k+2S)})\right]^2
  + \frac{\gamma^2   S}{(k-S)(k+2S)^2}=0. 
\label{exact_hyperbola}
\end{equation}
The line and hyperbola solutions are shown in
figs.~\ref{nonlinear_phase_space_flow} and \ref{fixed_points}. The
hyperbola intersects the transition line at the transition points.

The position of the transition line and points allows us to define two
critical switching rates.  When $S$ increases to the value
$S_{\rm low} = (\Koff-\Konc)/2$, the transition line first intersects
$(\sigma_R, \sigma_L)=(\frac{1}{2},\frac{1}{2})$. This allows the
H$_n$ phase to appear for $S_{\rm low}<S<S_{\rm high}$ (as discussed
below in sec.~\ref{sec:phase_diagram}). When
$S>S_{\rm high} = \Koff-\Konc$, the transition points appear. This
upper critical switching rate occurs when the transition line first
intersects $(\sigma_R, \sigma_L)=(\frac{1}{2},0)$ and
$(0,\frac{1}{2})$. This allows appearance of the LHLH phase (as
discussed below in sec.~\ref{sec:phase_diagram}).

The general case in equation \eqref{exact_2} is can be understood as a
exponential-like term (because the term $(\gamma+2S\phi)^{1+k/S}$
reduces to an exponential if $S \rightarrow 0$) times a hyperbola
term. The hyperbola has foci and, in general, two intersections on the
$\sigma_R = \sigma_L$ line. Multiplying by the exponential term does
not change these properties qualitatively, if it remains real. For the
special value
\begin{equation}
  C_{\rm LI}=\frac{\gamma S
    \frac{\gamma(k+S)}{k-S}^{2+k/S}}{(k+S)(k+2S)(k+3S)},\end{equation} 
the two intersections of the curves with the $\sigma_R = \sigma_L$
line become one intersection \footnote{This is analogous to the
  standard hyperbola equation $\frac{x^2}{a^2}-\frac{y^2}{b^2}=r^2$.
  The curve intersects the $x$-axis at $(\pm ar,0)$. These reduce to
  only one intersection when $r = 0$.}. The intersection is the LI. At
the Langmuir isotherm, the density on 
each lane is the Langmuir density set by binding/unbinding
equilibrium. As a result,
$\frac{d \sigma_L}{dx} = \frac{d \sigma_R}{dx} = 0$.  We note that in
the limit $S \rightarrow 0$, these curves merge with the $C=0$
curves discussed above.  In this limit, $C_{\rm LI} = 0$. 

\subsubsection{Domain walls}

In principle, the phase-plane density profile can be determined by
following the local velocity field, connecting the two points on the
plane that correspond to the lane end boundary conditions. Indeed, if
the boundary points both lie in the same quadrant of the plane, the
solution follows the local flow. However, in many cases the boundary
points lie in different quadrants, so that the boundary points cannot
be connected without crossing the lines $\sigma_R = 0$ or
$\sigma_L = 0$ where equations \eqref{flow_q+} or \eqref{flow_q-} are
ill defined. Then the solution will contain a domain wall at position
$x_w$ that must satisfy the matching condition
$\sigma(x_w-\epsilon)=-\sigma(x_w+\epsilon)$ (as discussed in
sec.~\ref{sec:meanfield}). In the phase plane, a domain wall therefore
appears as sign change of one of the densities
(fig.~\ref{example_flow}).

\subsubsection{Finite-size constraint}
\begin{figure}[t!]
\begin{centering}
\includegraphics[width=0.45 \textwidth]{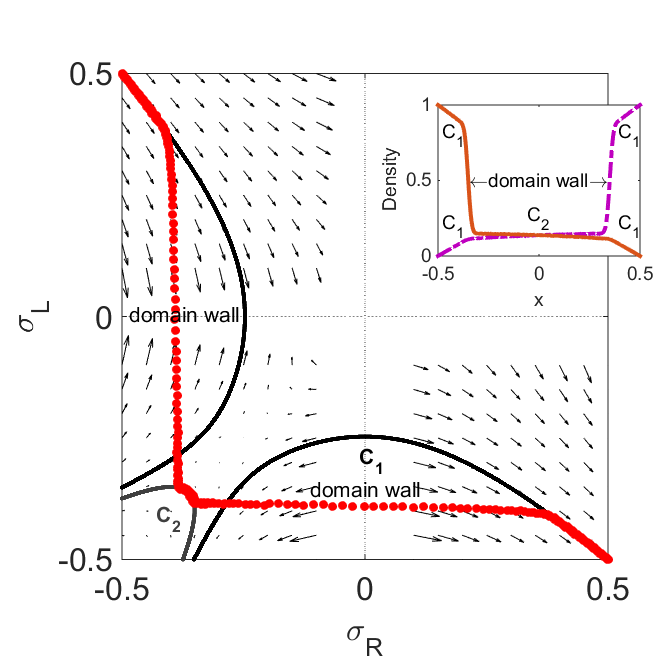} \includegraphics[width=0.45 \textwidth]{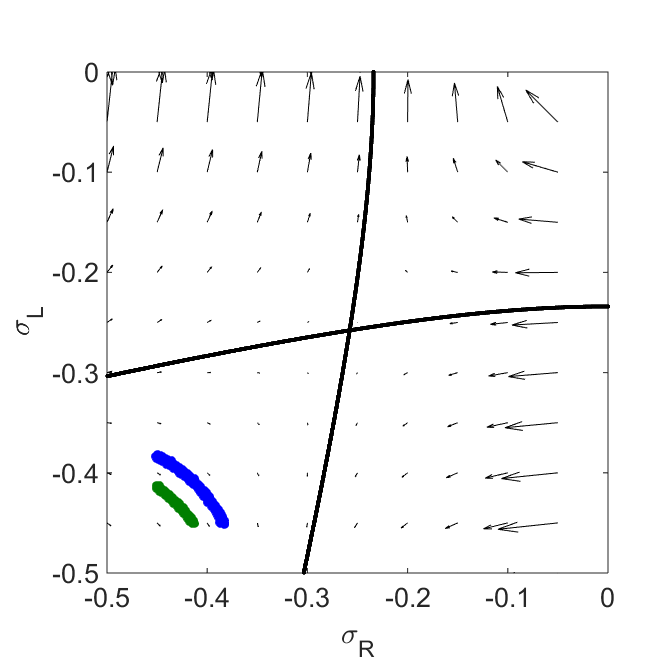}
\caption{Effects of domain walls and finite size on density
  profiles. Left, determination of domain wall positions in the phase
  plane trajectory and the density profile (inset). Red points show
  kMC simulation results. The boundary points are
  $(\sigma_R, \sigma_L) = (-\frac{1}{2}, \frac{1}{2})$ and
  $(\sigma_R, \sigma_L) = (\frac{1}{2}, -\frac{1}{2})$. The solution
  locally follows the flow, which cannot connect the boundary points
  without crossing the lines with $\sigma_R = 0$ and $\sigma_L = 0$
  where the velocity is ill-defined. Crossing these lines uses the
  matching condition to connect to another exact solution curve,
  introducing a domain wall. Inset shows the density profiles of
  $\rho_R$ (purple) and $\rho_L$ (brown). The sharp transitions in
  $\rho_R$ and $\rho_L$ indicate the domain walls. There are three
  regions which are separated by two domain walls: regions near ends
  follow $C_1$, and the region at the center is described by
  $C_2$. Right, finite-size constraint. The blue points are kMC
  simulation results for $N=1000$, and the green $N=500$. For a larger
  number of sites, the dimensionless motor speed is smaller, moving
  the solution closer to the LI. The switching rate is $0.44$\pers
  (left), $0.1$\pers (right), the bulk motor concentration $c = 200$
  nM, and the motor speed $5$ \mum\pers; other parameters are the
  reference values of table \ref{tab:units}. Arrows indicate vector
  field, which has the mathematical form in \eqref{flow_q+} and
  \eqref{flow_q-}.}
\label{example_flow}
\end{centering}
\end{figure}

The solutions are also affected by the finite-size constraint. If, for
example, each lane has 1000 sites, the correct trajectory should
connect the boundary points with exactly $1000$ sites. The number of
sites controls the effective magnitude of $dx$, and is therefore
analogous to time in the flow. Thus, the faster the effective flow,
the smaller the number of sites traversed in position space. At the
Langmuir isotherm, the number of sites can be infinite since this
point has zero flow velocity \footnote{Because the
  nondimensionalization depends on the motor speed, the connection
  between the phase-space effective velocity and the number of sites
  also depends on the motor speed.}. As the total number of sites
increases, the trajectory will approach closer to the LI, because this
point is the only one which can contain an infinite number of sites
(fig. \ref{example_flow}).  The finite-size constraint can prevent the
solution from exactly following the phase-space flow. As a result, the
boundary conditions are not always satisfied. This is discussed
further in sec.~\ref{sec:sym_phase_diagram}.

\section{Nonequilibrium phases for symmetric boundary conditions}
\label{sec:sym_phase_diagram}

The nonequilibrium steady-state solution of TASEP models sensitively
depends on the boundary conditions $\alpha$ and $1-\beta$
\cite{_nonequilibrium}. Because the flux $\rho(1-\rho)$ is maximized
for an occupancy of $\frac{1}{2}$, the phase with bulk density of
$\frac{1}{2}$ is called the maximum-current phase. The high-density
phase has bulk density $>\frac{1}{2}$ and the low-density phase has
bulk density $<\frac{1}{2}$. In the single-lane TASEP with LK, PFF
found a low density-high density coexistence phase and a Meissner
phase, but no maximum current phase \cite{parmeggiani_totally_2004}.

In our antiparallel two-lane model with binding and switching
kinetics, we find the same phases that appear in the single lane
case. In addition, we find a new four-phase coexistence low
density-high density-low density-high density (LHLH) phase, as
discussed below. In addition, the non-zero switching rate in our model
that couples the two lanes means that the central density is not only
attracted to the Langmuir isotherm and repelled from the maximum
current lines, but also attracted by the transition line. This
competition can cause either a local maximum or minimum of the density
at the overlap center ($x=0$).  If the total density
$\rho_R(x)+\rho_L(x)$ has a local maximum (minimum) at $x=0$, we
denote it a local maximum (minimum) phase. The occurrence of local
maxima/minima also occurs in the single lane case
\cite{parmeggiani_totally_2004}, though PFF didn't treat it as
separate feature of the phase since it has a less pronounced effect
there than in the antiparallel lane case where the overall density is
the sum of the two single-lane densities.

Here we focus on the case of symmetric boundary conditions with
$\alpha_R=\alpha_L = \alpha$ and $\beta_R=\beta_L = \beta$.  We first
discuss the stability of the boundary conditions and propeties of the
central density. Then, we describe each phase and how we determine the
phase boundaries.  We focus on the case LI $< \frac{1}{2}$. Because
the system has particle-hole symmetry, the case LI $> \frac{1}{2}$ can
be understood by the transformation $\rho \rightarrow 1-\rho$.

\begin{figure}[t!]
\begin{centering}
\includegraphics[width=0.45 \textwidth]{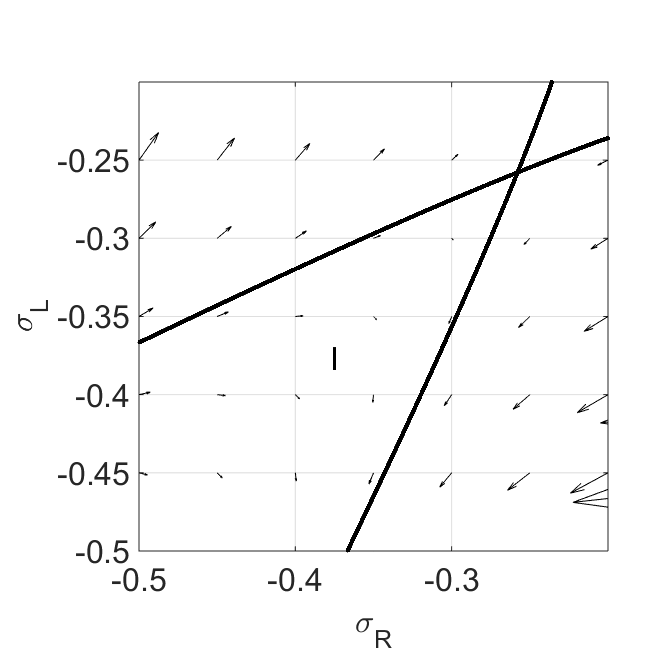}
\includegraphics[width=0.45 \textwidth]{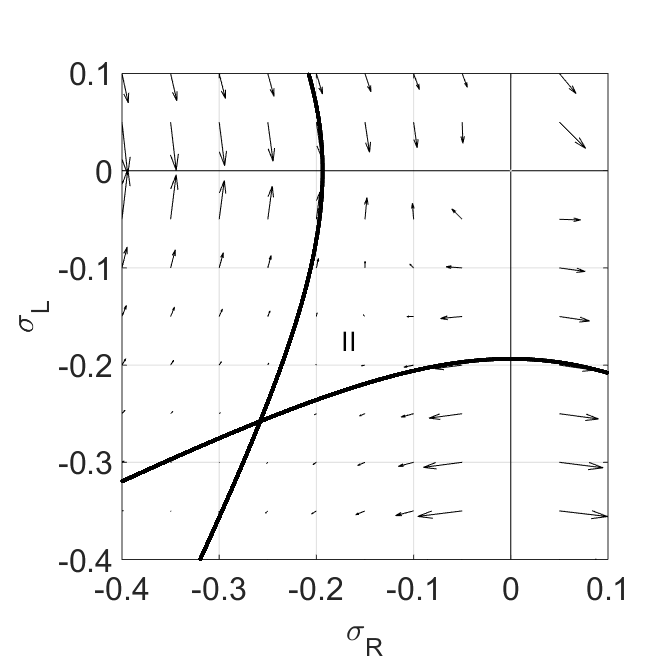}
\includegraphics[width=0.45 \textwidth]{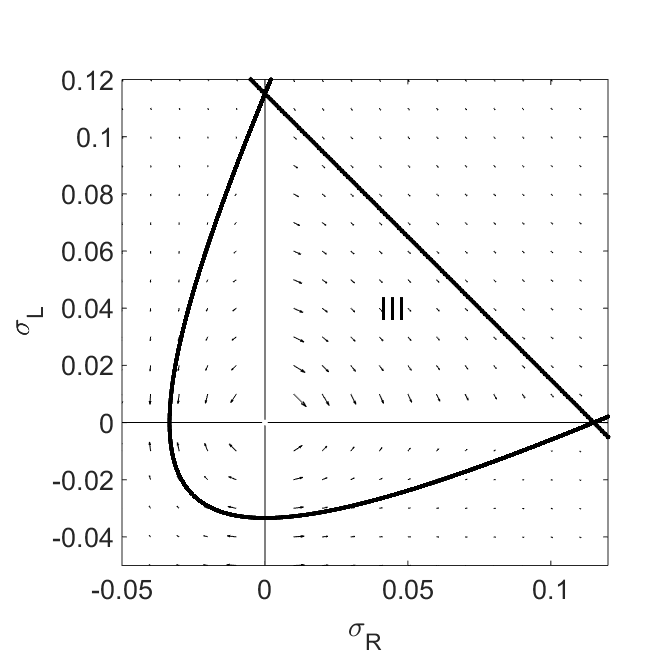}
\includegraphics[width=0.45 \textwidth]{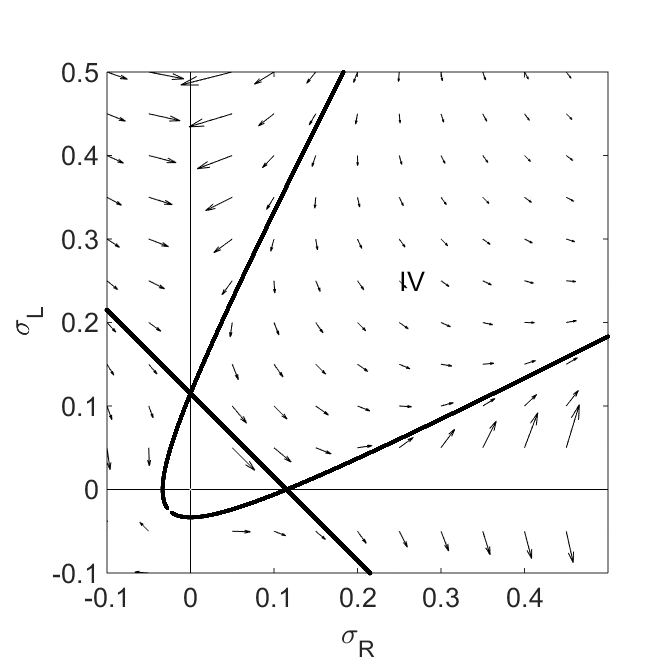}
\caption{Illustration of the 4 regions of the central density, as
  defined in the text.  The switching rate is $0.5$\pers, the bulk
  motor concentration $c = 200$ nM, and the motor speed is $5$
  \mum\pers; other parameters are the reference values of table
  \ref{tab:units}. Arrows indicate vector field, which has the mathematical form in \eqref{flow_q+} and \eqref{flow_q-}.}
\label{Regions}
\end{centering}
\end{figure}

\subsection{Domain wall motion}

In the steady-state TASEP, the boundary conditions are not always
satisfied at the boundary sites (or continuously approaching the
boundary).  The stability of the boundary density values was
determined for the single-lane TASEP by Kolomeisky et
al. \cite{kolomeisky_phase_1998}, who worked out the speed at which a
domain wall moves.  When $\alpha$ and $\beta < \frac{1}{2}$, the
domain wall velocity is $V = \frac{j_r-j_l}{\rho_r-\rho_l}$, where
$j_{r,l}$ denotes the current at the right and left boundaries. If we
take $\rho_l=\alpha$, $j_l= \alpha(1-\alpha)$, $\rho_r=\beta$, and
$j_r=\beta(1-\beta)$, the domain wall velocity is $\beta-\alpha$.
Therefore if $\beta > \alpha$ the domain wall moves to the right end
of the system and the right boundary condition is not satisfied, while
if $\beta < \alpha$ the left boundary condition is not satisfied.

A similar relation can be determined for matching a high-density
region to a maximum-current region with a domain wall
\cite{kolomeisky_phase_1998}. Suppose $\alpha > \frac{1}{2}$ and
$\beta < \frac{1}{2}$, but a maximum-current phase appears on the left
so that $j_l$ becomes $\frac{1}{4}$.  The domain wall velocity becomes
$V = \frac{j_r-j_l}{\rho_r-\rho_l} =
\frac{\beta(1-\beta)-\frac{1}{4}}{1-\beta-\frac{1}{2}}=\beta-\frac{1}{2} < 0$,
which gives an unstable left boundary condition.  Similar behavior
occurs if $\alpha < \frac{1}{2}$ and $\beta > \frac{1}{2}$. 

These relations no longer strictly hold when binding kinetics or
switching between multiple lanes are added to the model. However, they
are a valuable starting point to gain intuition about the stability of
domain walls due to the TASEP.

\subsection{Properties of the central density}
\label{sec:maximin}

When the boundary conditions are symmetric, the total density is
symmetric about $x=0$.  Therefore, $\rho_R(0)=\rho_L(0)$ and the
density in the center of the system must lie on the
$\sigma_R = \sigma_L$ line. The $\sigma_R = \sigma_L$ line can be
separated into four regions, which correspond to four different
possible behaviors of the central density
(fig. \ref{nonlinear_phase_space_flow} and \ref{Regions}).

Region I occurs where $\sigma_R = \sigma_L$ and the density is less
than the Langmuir isotherm. In this region, the flow makes the density
approach the LI. At the isotherm, the rates of change of both
$\sigma_R$ and $\sigma_L$ are zero, and as the flow approaches the LI,
these rates of change decrease. Therefore, the density remains less
than or equal to the Langmuir isotherm. This makes the central density
a local maximum, which we denote a local maximum phase.  Region II
occurs where $\sigma_R = \sigma_L$ and the density is greater than the
Langmuir isotherm, but less than 0. In this region,
the flow makes the density approach the LI, decreasing the
density. The density remains greater than or equal to the Langmuir
isotherm. This makes the central density a local minimum, which we
denote a local minimum phase.

Regions III and IV occur where $\sigma_R = \sigma_L$ and the density
is greater than 0 but less than (region III) or greater than (region
IV) the transition line. The transition line occurs where the effects
of switching and binding kinetics balance. In region III, binding
kinetics are more important, while in region IV, switching kinetics
are more important.  In both of these regions, the central density is
$>\frac{1}{2}$ on a single lane. Because the density is greater than
the LI, the mean-field binding and unbinding terms in equations
\eqref{eq:cont1} and \eqref{eq:cont2} are net negative. As a result,
the flux term
$-(2\rho_R-1) \frac{\partial \rho_R}{\partial x} = \frac{\partial
}{\partial x} (\rho_R(1-\rho_R)) =
(\rho_R(x_{out})(1-\rho_R(x_{out})))-(\rho_R(x_{in})(1-\rho_R(x_{in})))$
becomes negative. Therefore, the flux decreases as the density
increases. Another way to see this is to note that the flux has a
maximum for $0\le \rho\le 1$ when $\rho = \frac{1}{2}$. For
$\rho < \frac{1}{2}$, the flux increases as density increases, while
if $\rho > \frac{1}{2}$, the flux decreases as density increases.  The
net effect is to cause the density to increase approaching the
$\sigma_R = \sigma_L$ line in region III. Therefore the central
density has a local maximum in region III.  However, in region IV,
$\sigma_R$ and $\sigma_L$ are larger, causing the switching terms to
contribute significantly to the flow. This gives a positive
contribution in equations \eqref{eq:cont1} and
\eqref{eq:cont2}. Therefore, the flow changes the sign in region IV
(compared with region III). The central density has a local minimum in
region IV.

\subsection{Nonequilibrium phases}
\begin{figure}[t!]
\begin{centering}
\includegraphics[width=0.9 \textwidth]{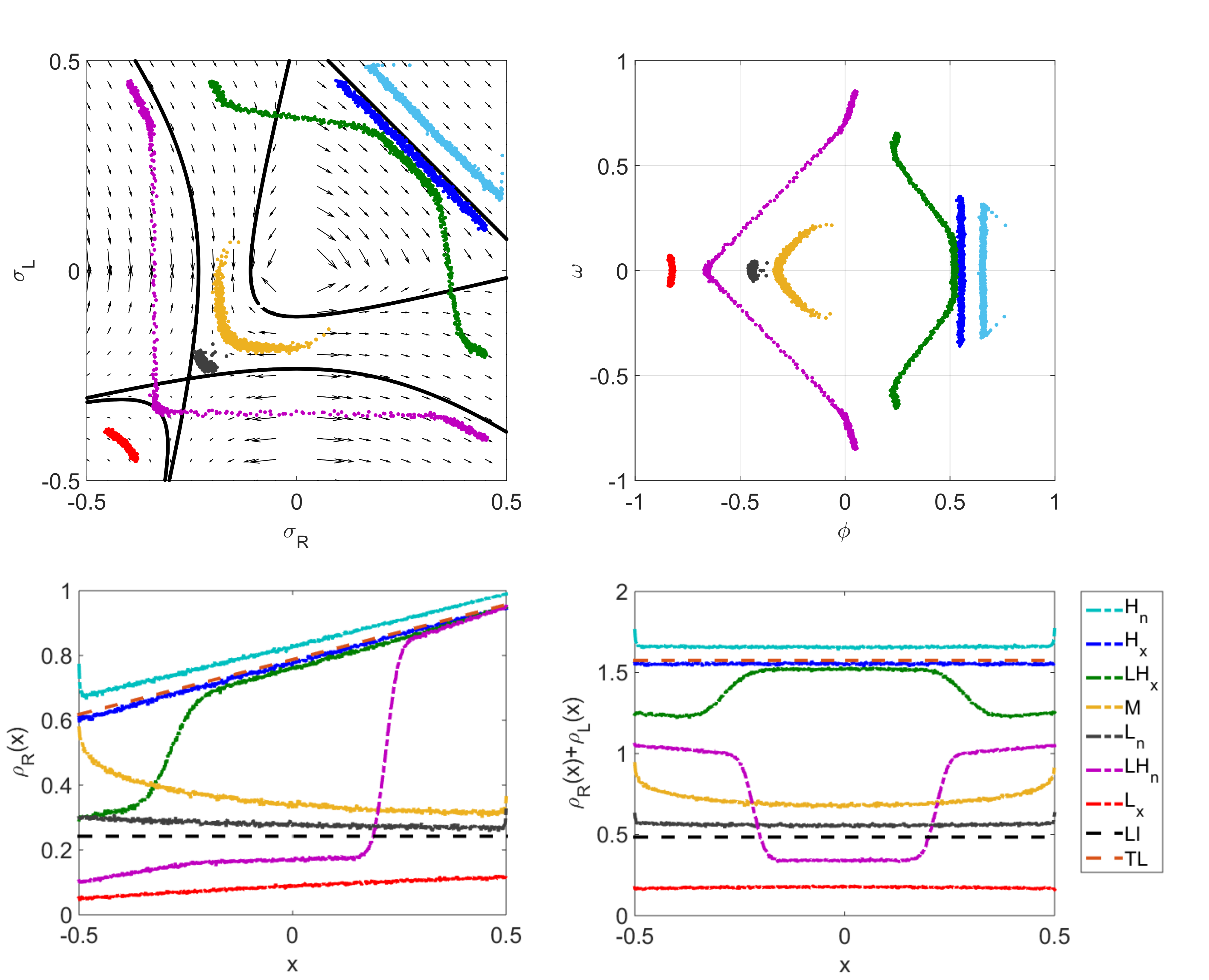}
%\includegraphics[width=0.45 \textwidth]{nonlinear_flow_f_6_2}
%\includegraphics[width=0.45 \textwidth]{nonlinear_flow_f_6_2_phi_omega}
%\hfill\flushleft
%\includegraphics[width=0.45 \textwidth]{example_of_phases}
%\includegraphics[width=0.45 \textwidth]{example_of_phases_2}
\caption{Examples of the nonlinear phases for low switching rate.
  Upper left: trajectories in the $\sigma_R$-$\sigma_L$ plane. Upper
  right: trajectories in the $\phi$-$\omega$ plane, where
  $\phi = \sigma_R+\sigma_L$ and $\omega = \sigma_R-\sigma_L$,
  illustrating the local maxima and minima of the central
  density. Lower left: the density in lane R, $\rho_R(x)$. The black
  dashed line indicates the LI in position space, and the brown dashed
  line is the transition line in position space. Lower right: the
  total density as a function of position. The black dashed line
  indicates the LI (its value is $2\rho_0$, since it is
  $\rho_R+\rho_L$) in position space, and the brown dashed line is the
  transition line in position space (its value is
  $-\frac{\gamma}{2S}+1$, since it is $\rho_R+\rho_L$). The boundary
  conditions are $(\alpha, 1-\beta) = (0.05, 0.1)$, red,
  $(0.1, 0.95)$, purple, $(0.3, 0.4)$, grey, $(0.6, 0.4)$, yellow
  $(0.3, 0.95)$, green, $(0.6, 0.95)$, blue, and $(0.95, 0.99)$,
  cyan. The switching rate is $0.1$\pers, the bulk motor concentration
  $c = 200$ nM, and the motor speed $5$ \mum\pers; other parameters
  are the reference values of table \ref{tab:units}. Arrows in the
  upper left figure indicate vector field, which has the mathematical
  form in \eqref{flow_q+} and \eqref{flow_q-}.}
\label{nonlinear_phase_space_flow_3}
\end{centering}
\end{figure}

\begin{figure}[t!]
\begin{centering}
\includegraphics[width=0.9 \textwidth]{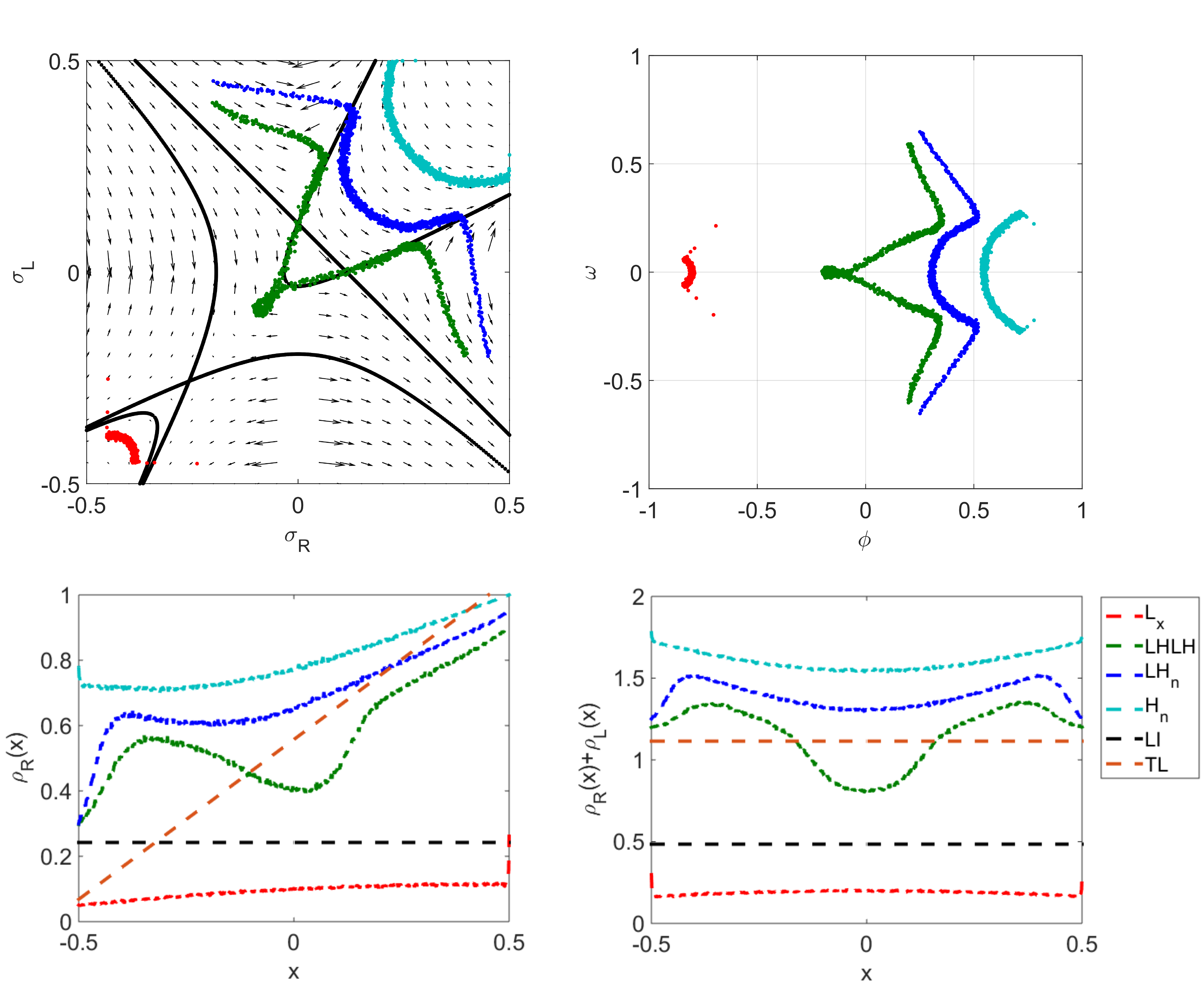}
\caption{Examples of the nonlinear phases for high switching rate.
  Upper left: trajectories in the $\sigma_R$-$\sigma_L$ plane. Upper
  right: trajectories in the $\phi$-$\omega$ plane, where
  $\phi = \sigma_R+\sigma_L$ and $\omega = \sigma_R-\sigma_L$,
  illustrating the local maxima and minima of the central
  density. Lower left: the density in lane R, $\rho_R(x)$. The black
  dashed line indicates the LI in position space, and the brown dashed
  line is the transition line in position space. Lower right: the
  total density as a function of position. The black dashed line
  indicates the LI (its value is $2\rho_0$, since it is
  $\rho_R+\rho_L$) in position space, and the brown dashed line is the
  transition line in position space (its value is
  $-\frac{\gamma}{2S}+1$, since it is $\rho_R+\rho_L$). The boundary
  conditions are $(\alpha, 1-\beta) = (0.05, 0.6)$, red, $(0.3, 0.9)$,
  green, $(0.3, 0.95)$ blue, and $(0.9, 1.0)$, cyan. The switching
  rate is $0.5$\pers, the bulk motor concentration $c = 200$ nM, and
  the motor speed $5$ \mum\pers; other parameters are the reference
  values of table \ref{tab:units}. Arrows in the upper left figure
  indicate vector field, which has the mathematical form in
  \eqref{flow_q+} and \eqref{flow_q-}.}
\label{nonlinear_phase_space_flow_4}
\end{centering}
\end{figure}

Here we discuss the nonequilibrium phases that occur in our model, as
illustrated in figs.~\ref{nonlinear_phase_space_flow_3},
\ref{nonlinear_phase_space_flow_4}. There are 5 possible phases: low
density, high density, low density-high density, low density-high
density-low density-high density, and Meissner.  As mentioned above,
we focus on the case with LI $< \frac{1}{2}$.

\begin{figure}[t!]
\begin{centering}
\includegraphics[width=0.45 \textwidth]{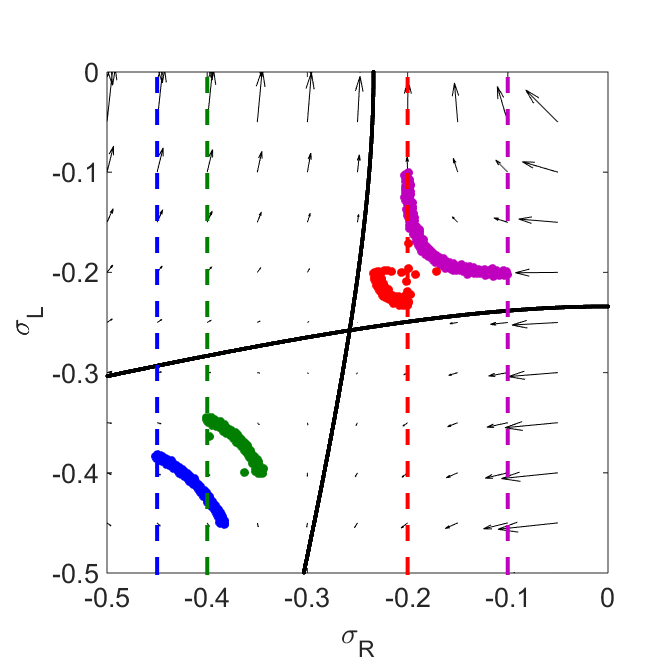}
\includegraphics[width=0.45 \textwidth]{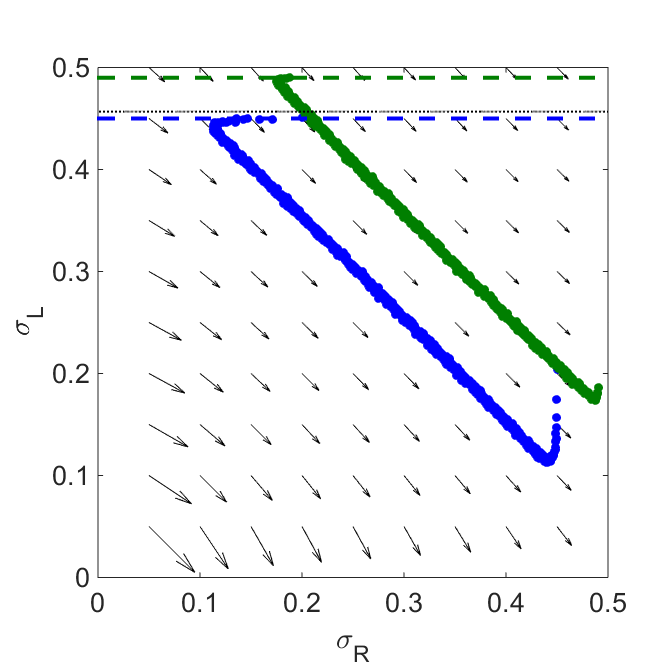}
\caption{Example phase-plane trajectories of L and H phases. Left: L
  phases.  The starting and ending points are
  $(\sigma_R,\sigma_L) = (-0.45, -0.4)$ and $(-0.4, -0.45)$, blue,
  $(-0.4, -0.45)$ and $(-0.45, -0.4)$, green, $(-0.2, -0.1)$ and
  $(-0.1, -0.2)$, red, and $(-0.1, -0.2)$ and $(-0.2, -0.1)$,
  purple. The dashed lines show that the L phase obeys the left
  boundary conditions, where the positions are $\sigma_R = -0.45$,
  blue, $\sigma_R = -0.4$, green, $\sigma_R = -0.2$, red, and
  $\sigma_R = -0.1$, purple.  Right: H phases. The starting and ending
  points are $(\sigma_R,\sigma_L) = (0.3, 0.45)$ and $(0.45, 0.3)$,
  blue, $(0.2, 0.49)$ and $(0.49, 0.2)$, green. The dashed lines show
  that the H phase obeys the right boundary conditions, where the
  positions are $\sigma_L = 0.45$, blue, and $\sigma_L = 0.49$, green.
  The black dashed line
  $\sigma_L = \rho_c-\frac{1}{2} \approx 0.4567$, the critical density
  $\rho_c$ (see text) of the local minimum and local maximum
  phases. The switching rate is $0.1$\pers, bulk motor concentration
  $c = 200$ nM, and motor speed $5$ \mum\pers; other parameters are
  the reference values of table \ref{tab:units}. Arrows indicate vector field, which has the mathematical form in \eqref{flow_q+} and \eqref{flow_q-}.}
\label{LD_HD_boundary_stability}
\end{centering}
\end{figure}

\textbf{Low density (L)}: The density in each lane remains
$<\frac{1}{2}$.  The L phase occurs when $\alpha < \frac{1}{2}$ and
the right boundary condition cannot be satisfied.  According to the
behavior of the local maxima and minima derived above in
sec.~\ref{sec:maximin} above, if $\alpha < $ LI, the central density
is a local maximum; if $\alpha > $ LI, the central density is a local
minimum (fig. \ref{LD_HD_boundary_stability} right).

\textbf{High density (H):} The density in each lane remains
$>\frac{1}{2}$. The H phase occurs when $\alpha > \frac{1}{2}$ and
$\beta<\frac{1}{2}$ so that the left boundary condition cannot be
satisfied. There is a critical concentration we denote $\rho_c$ which
corresponds to the density where the transition line crosses the
$\sigma_R=\sigma_L$ line.  If $1-\beta < \rho_c$, the central density
is a local maximum, while if $1-\beta > \rho_c$, the central density
is a local minimum (fig. \ref{LD_HD_boundary_stability} left).

\textbf{Low density-high density coexistence (LH):} In this phase, the
steady-state density on each lane has both low-density and
high-density regions. A domain wall occurs where the low- and
high-density phases meet; we call the length of the high-density
region the boundary-layer length \cite{kuan_motor_2015}. We discuss
how the domain wall position is determined in more detail in
sec.~\ref{sec:phase_diagram}.  The boundary layer length determines
whether the overlap shows greater motor accumulation at the center or
at the ends: if the boundary layer length is greater than half the
overlap length, the overall density is higher at the overlap center
(since we set our LI $ < \frac{1}{2}$). We note that this condition is
distinct from whether or not a local maximum or minimum occurs at the
overlap centers, as discussed above.  The local extremum is identified
using the derivative $d (\sigma_R+\sigma_L)/dx$ at the overlap center.

%%mb: editing this paragraph %%
\textbf{Low density-high density-low density-high density (LHLH):} For
sufficiently high switching rate, a multi-phase coexistence region can
appear in the LH region (fig. ~\ref{nonlinear_phase_space_flow_4}).
To understand why the LHLH phase occurs, note that the transition line
(fig. \ref{nonlinear_phase_space_flow} left) divides the phase plane
into two regions. If a boundary point is above the transition line in
the upper left quadrant, the flow does not reach the $\sigma_L = 0$
line, where a jump into region II is possible (fig. \ref{Regions}) in
order to connect the flow to the other boundary condition
\footnote{The reason why we mention region II is for two reasons:
  first, it is typically not possible to jump to region I, because of
  the matching condition
  $\sigma(x_w-\epsilon) = -\sigma(x_w+\epsilon)$.  To jump from the
  upper left quadrant to the lower left quadrant, $\sigma_R$ should be
  the same. However, the region which is above the transition line and
  in the upper left quadrant might have higher $\sigma_R$ value than
  the highest possible $\sigma_R$ value in region I.  Second, if
  jumping to region III and IV, the trajectory cannot be completed
  with the correct total number of sites.}. Thus, the only allowed
domain wall involves a jump to the upper right quadrant (outside of
the hyperbola), followed by further motion along the flow field. The
solution then jumps to region II, because the flow outside the
hyperbola will not cross the line $\sigma_R = \sigma_L$.  These
multiple jumps cause multiple domain walls to appear. We note that the
LHLH phase is reminiscent of LD-BP-HD multi-phase coexistence found by
Pierobon et al.~\cite{pierobon_bottleneckinduced_2006}, which arises
from a point defect on a single lane.

\textbf{Meissner (M):} In the pure TASEP, the maximum current phase
occurs when the bulk density profile is independent of the boundary
conditions \cite{_nonequilibrium}.  The analogous phase in the TASEP
with LK is the Meissner phase \cite{parmeggiani_totally_2004}. Neither
of the boundary conditions is satisfied in this phase.

\begin{figure}[t!]
\begin{centering}
\includegraphics[width=0.45 \textwidth]{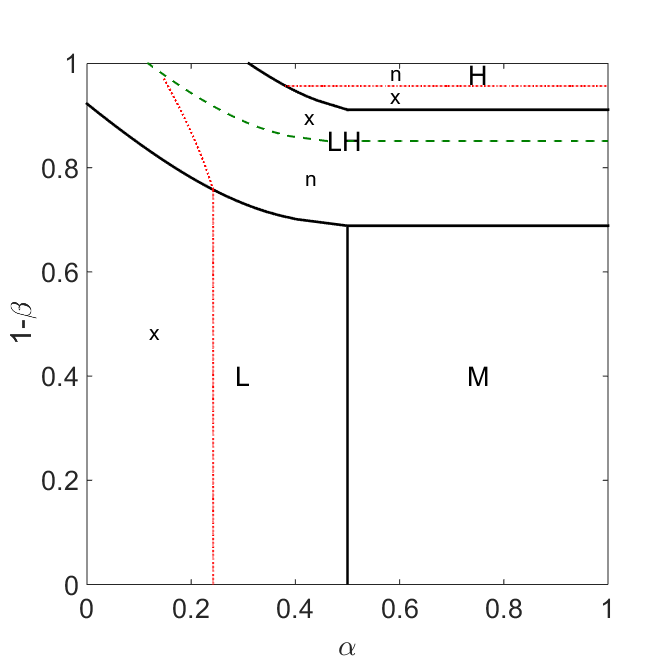}
\includegraphics[width=0.45 \textwidth]{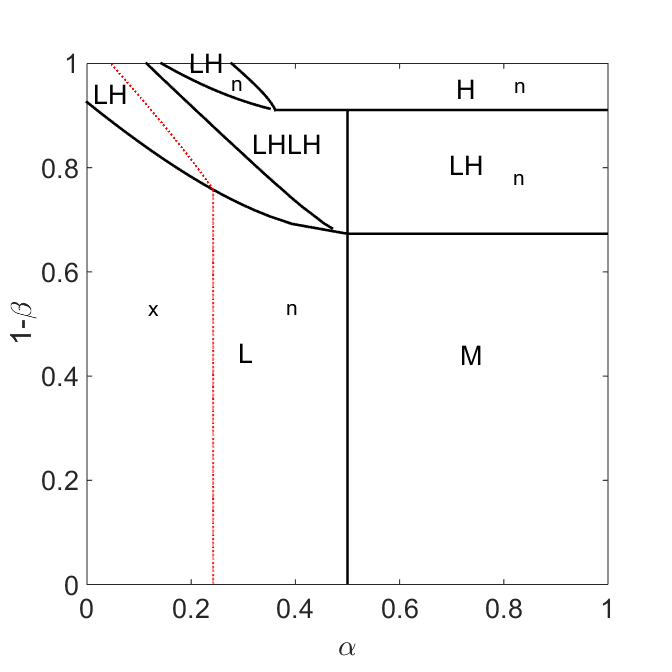}
\caption{The phase diagrams for low (left, $0.1$\pers) and high
  (right, $0.5$\pers) switching rate. The phases are L, low density,
  M, Meissner, H, high density, LH, low density-high density
  coexistence, and LHLH, low density-high density-low density-high
  density coexistence. The green dashed line indicates where the
  domain wall position $x_w = 0$, which separates regions with a local
  maximum of the central density (x) or local minimum (n). The red
  dashed lines indicate boundaries between local maximum and minimum
  phases. For low switching rate, the left parts of the L and LH
  phases are local maximum phases, and the right are local minimum
  phases. The upper part in the H phase is local minimum phase. For
  high switching rate, the left parts of the L and LH phases are local
  maximum phases, and the right parts are local minimum phases. The
  bulk motor concentration is $c = 200$ nM, and the motor speed $5$
  \mum\pers; other parameters are the reference values of table
  \ref{tab:units}.}
\label{phase_diagram}
\end{centering}
\end{figure}

\section{Phase diagram for symmetric boundary conditions}
\label{sec:phase_diagram}

Figure \ref{phase_diagram} shows typical phase diagrams illustrating
the regions where the five phases (L, H, M, LH, and LHLH) appear as a
function of the boundary conditions $\alpha$ and $1-\beta$. These
boundary motor densities define the boundary points in the phase plane
(fig. \ref{nonlinear_phase_space_flow}). To determine the phase
diagram, it is convenient to determine trajectories on the phase
plane. The phase regions are the collection of all the boundary points
which show the same physical behavior
(fig. \ref{nonlinear_phase_space_flow_3} and
\ref{nonlinear_phase_space_flow_4}). As noted above, we study LI
$< \frac{1}{2}$. The phase boundaries are determined as follows:

\textbf{Boundary between L and M phases:} The phase boundary between L
and M occurs where $\alpha = \frac{1}{2}$, because the boundary
conditions cannot be satisfied when $\alpha > \frac{1}{2}$ and
$\beta > \frac{1}{2}$. The phase in which none of the boundary
conditions are satisfied defines the Meissner phase
\cite{parmeggiani_totally_2004}.

\textbf{Boundaries of the LH phase:} The low density-high density
coexistence phase contains a domain wall, at which the density changes
discontinuously but the flux $\rho(1-\rho)$ is remains continuous
\cite{parmeggiani_totally_2004}. Across the domain wall, the matching
condition $\sigma(x_w-\epsilon) = -\sigma(x_w+\epsilon)$ must be
satisfied. The boundaries of the LH phase occur when the domain wall
position $x_w$ moves outside the lane, that is, when
$|x_w|>\frac{1}{2}$.

\begin{figure}[t!]
\begin{centering}
\includegraphics[width=0.45 \textwidth]{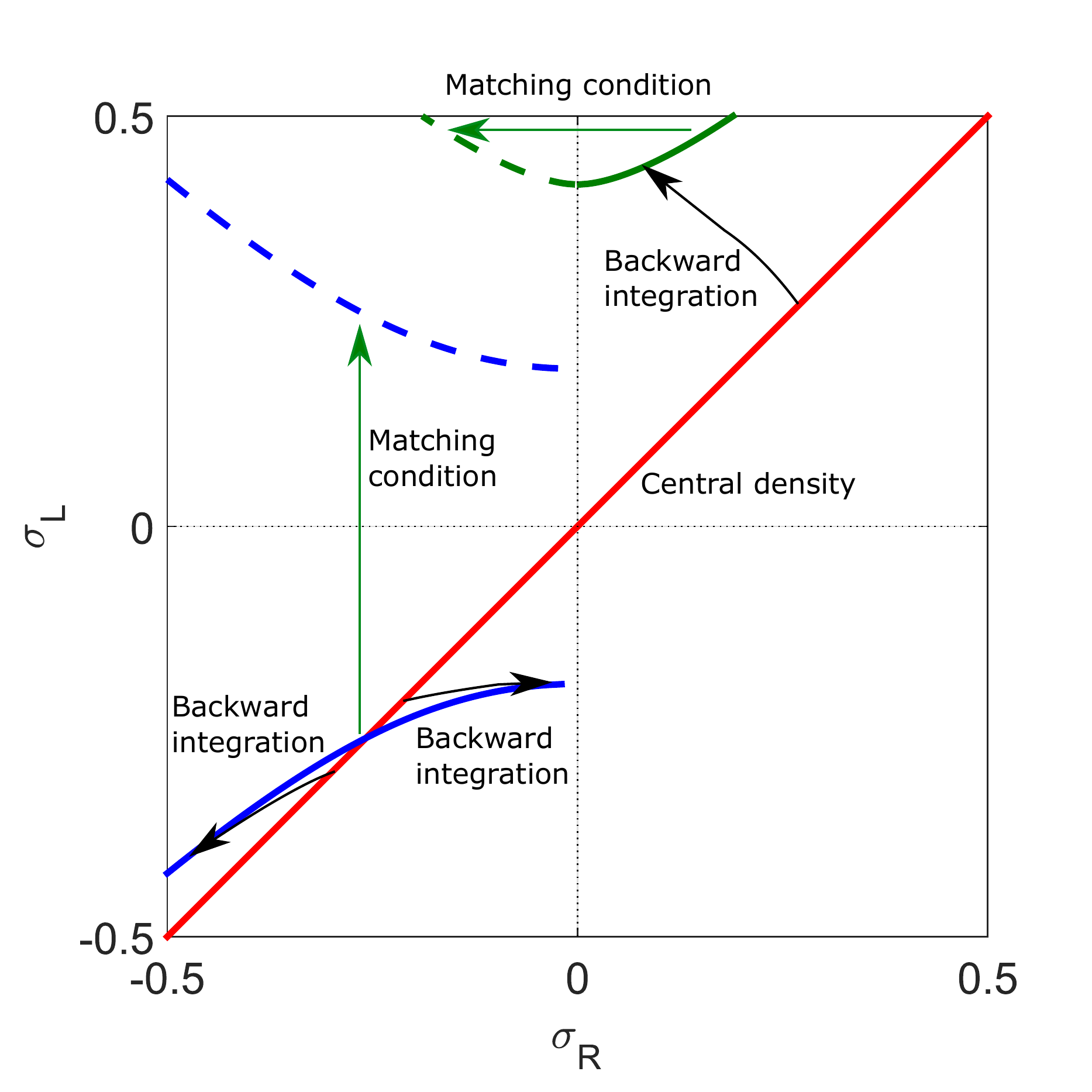}
\caption{Illustration of calculation of the LH phase boundary. The red
  line indicates the central density, which lies on the
  $\sigma_R = \sigma_L$ line for symmetric boundary
  conditions. Backward integration from this line (black arrows) to
  $x=-\frac{1}{2}$ gives the thick blue and green, where the blue line
  corresponds to $\sigma < 0$ and green line $\sigma > 0$. Applying
  the matching condition to $\sigma_L$ and $\sigma_R$ (green arrows)
  gives the dashed blue and green lines, respectively. These are the
  phase boundaries of the LH phase.  The switching rate is $0.1$\pers,
  bulk motor concentration $c = 200$ nM, and motor speed $5$
  \mum\pers; other parameters are the reference values of table
  \ref{tab:units}.}
\label{LH_phase_boundary}
\end{centering}
\end{figure}

To determine the domain wall position, we integrate the density
profile back from the center ($x=0$) to $x = -\frac{1}{2}$, thereby
determining $\sigma_R(x=-\frac{1}{2})$. Because system is symmetric,
the central density $\sigma_{R,L}(x=0)$ must lie on the line
$\sigma_R = \sigma_L$ (fig. \ref{nonlinear_phase_space_flow}). By
integrating the density to $x=-\frac{1}{2}$, we can map the line of
slope 1 to the set of points $\sigma_R(x=-\frac{1}{2})$
(fig.~\ref{LH_phase_boundary}, black arrows to blue solid lines).  If
we apply the matching condition to this set of points to jump them to
positive values of $\sigma_L$, we have found the set of points for
which the domain wall occurs at $x_w = \frac{1}{2}$. This is
equivalent to an L phase that extends to the right end of the lane
(fig. \ref{LH_phase_boundary}, green arrow to blue dashed lines).
This defines the phase boundary of the LH phase
(fig. \ref{LH_phase_boundary}, blue dashed line).  Using the same
analysis, we draw another phase boundary of the LH phase when the
domain wall position is at $x_w = -\frac{1}{2}$, which is equivalent
to the boundary for which the high-density phase extends to the left
end of the system (fig. \ref{LH_phase_boundary}, green dashed line).

In the upper right quadrant of the phase plane ($\sigma_{R, L}>0$),
the left boundary condition cannot be satisfied, so the LH phase
boundaries are the horizontal lines which define the upper boundary of
the M phase. If the domain wall position is greater than
$\frac{1}{2}$, the right boundary condition is no longer satisfied,
and vice versa for $x_w<-\frac{1}{2}$ case. Thus, the region under the
lower boundary of the LH phase and $\alpha > \frac{1}{2}$ does not
satisfy the right boundary condition. In addition,
$\alpha > \frac{1}{2}$ means that the left boundary condition is not
satisfied. Therefore, this region is the Meissner phase.

\begin{figure}[t!]
\begin{centering}
\includegraphics[width=0.45 \textwidth]{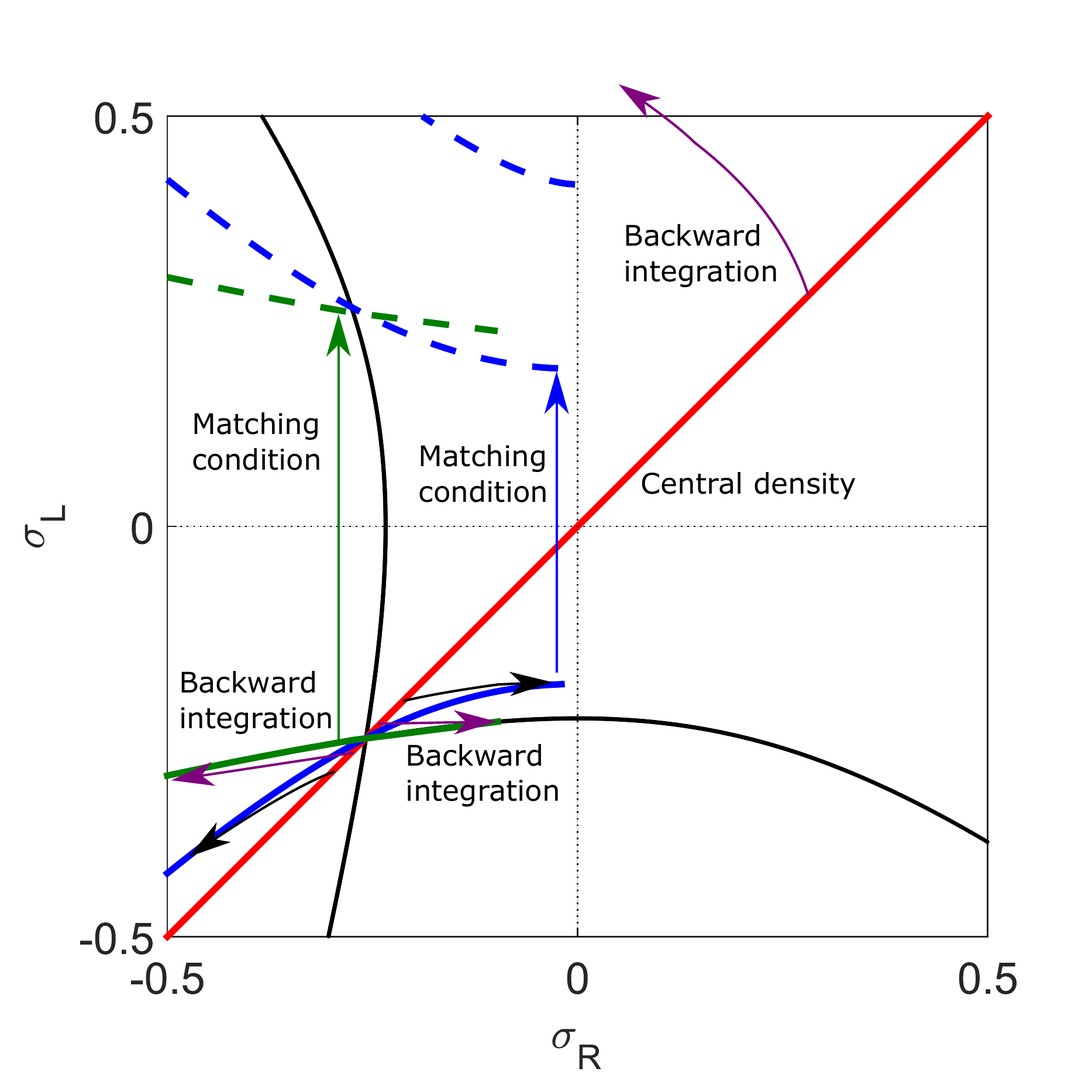}
\caption{Illustration of the effects of changing motor speed on the LH
  phase boundary. High motor speed(5 \mum\pers) shown in blue, low
  motor speed $0.5$ \mum\pers shown in green. The red line indicates
  the central density, which lies on the $\sigma_R = \sigma_L$ line
  for symmetric boundary conditions. Backward integration from this
  line (black and purple arrows) to $x=-\frac{1}{2}$ gives the thick
  blue and green lines. Applying the matching condition to $\sigma_L$
  and $\sigma_R$ (blue and green arrows) gives the dashed blue and
  green lines. These are the phase boundaries of the LH phase. The
  solid black lines are trajectories which pass through the Langmuir
  isotherm. For lower motor speed, backward-integrated line moves
  closer to the black line.  The phase boundary boundary of the LH and
  H phase is outside of the lanes. The switching rate is $0.1$\pers,
  the bulk motor concentration $c = 200$ nM; other parameters are the
  reference values of table \ref{tab:units}.}
\label{LH_phase_boundary_hl}
\end{centering}
\end{figure}

Since the phase boundaries depend on assuming that the points
$\sigma_{R,L}(x=0)$ lie on the $\sigma_R = \sigma_L$ line, the width
of the phases and the shape of the phase boundaries depend on the
number of sites in the lanes and the motor speed. The faster the speed
or the lower the number of sites, the smaller the dimensionless values
of $\Konc$, $\Koff$, and $S$. This leads to a smaller magnitude of the
phase plane flow velocity.  Since the integration of the densities
from $x=0$ to $x=- \frac{1}{2}$ is inversely proportional to the flow
velocity $\frac{\partial \sigma_{R,L}}{\partial x}$, decreases in
motor speed make the phase boundary lines closer to the line
$\sigma_R = \sigma_L$. However, the line dividing the L and LH phases
always passes through the Langmuir isotherm
$\sigma_R=\rho_0-\frac{1}{2}$, $\sigma_L= -\rho_0+\frac{1}{2}$, since
the phase plane flow velocity is zero at the Langmuir isotherm.
Similarly, slower motor speed or higher number of sites in the overlap
makes the set of points obtained by integrating backward move closer
to the trajectory line that passes through the Langmuir isotherm
(fig. \ref{LH_phase_boundary_hl}).

Once the domain wall position decreases to $x_w<0$, the longer length
of the high-density region makes the overall central density greater
than the end density. This determines whether the center of the lanes
has a local maximum or local minimum; these two cases are
distinguished by the thick dashed line in the phase diagram
(fig.~\ref{phase_diagram} left). If $x_w=0$, then the central density
must lie on the line $\sigma_R = -\sigma_L$, because then by the
matching condition the density can jump to the line
$\sigma_R = \sigma_L$ and then jump to $\sigma_R = -\sigma_L$.
Therefore, the set of starting points that correspond to domain wall
positions with $x_w=0$ is calculated by integrating backwards from the
$\sigma_R = -\sigma_L$ line to determine
$\sigma_{R,L}(x=-\frac{1}{2})$.

The weak dashed line in the LH phase indicates whether a local maximum
or minimum occurs (fig. \ref{phase_diagram}). If the domain wall is in
region I, the central density is a local maximum, while if domain wall
is in region II, the central density is a local minimum.  By
integrating backwards from $\sigma_R=\rho_0-\frac{1}{2}$,
$\sigma_L= -\rho_0+\frac{1}{2},$ (the point at which the domain wall
position overlaps with the Langmuir isotherm), we can determine the
dividing line between the local maximum/minimum regions \footnote{We
  note that the H phase has a local maximum or minimum due to the
  transition line. However, there is no local maximum and minimum due
  to the transition line in the LH phase, because the boundary of the
  local maximum and minimum due to the transition line in the LH phase
  is outside of the phase boundary. In other words, this case only
  occurs when the domain wall position is outside of the lane.}.

\begin{figure}[t!]
\begin{centering}
\includegraphics[width=0.45 \textwidth]{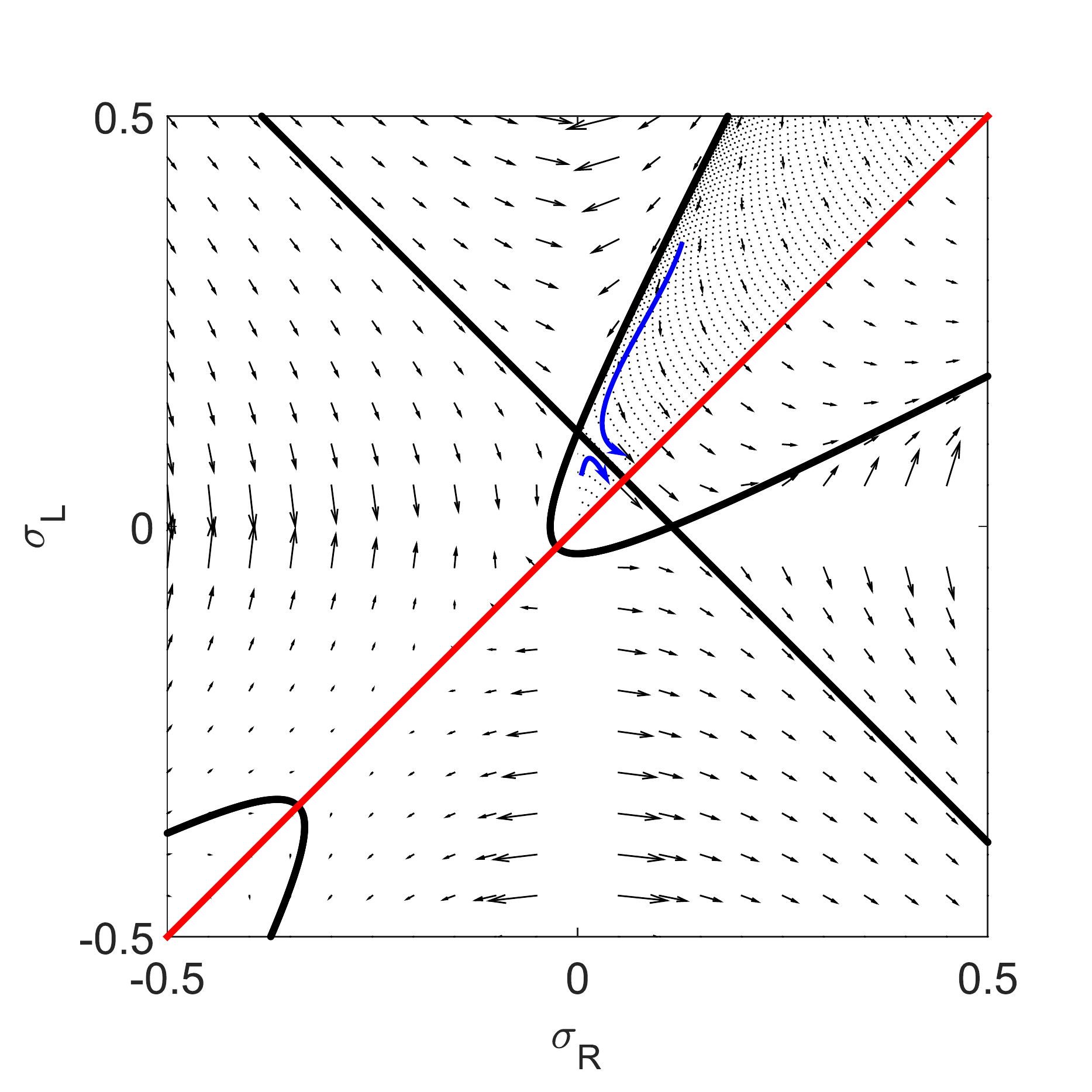}
\caption{Illustration of calculation of the LHLH phase boundary. The
  red line indicates the central density, which lies on the
  $\sigma_R = \sigma_L$ line for symmetric boundary conditions. The
  thick black lines are the line and hyperbola determined from the
  equation \eqref{exact} with $C = 0$. The dashed black lines are trajectories
  determined by backward integration the from $\sigma_R=\sigma_L$
  line. The blue curves show the flow directions. The switching rate
  is $0.5$\pers, bulk motor concentration $c = 200$ nM, and motor
  speed $5$ \mum\pers; other parameters are the reference values of
  table \ref{tab:units}. Black arrows indicate vector field, which has the mathematical form in \eqref{flow_q+} and \eqref{flow_q-}.}
\label{LHLH_phase_boundary}
\end{centering}
\end{figure}

\begin{figure}[t!]
\begin{centering}
\includegraphics[width=0.45 \textwidth]{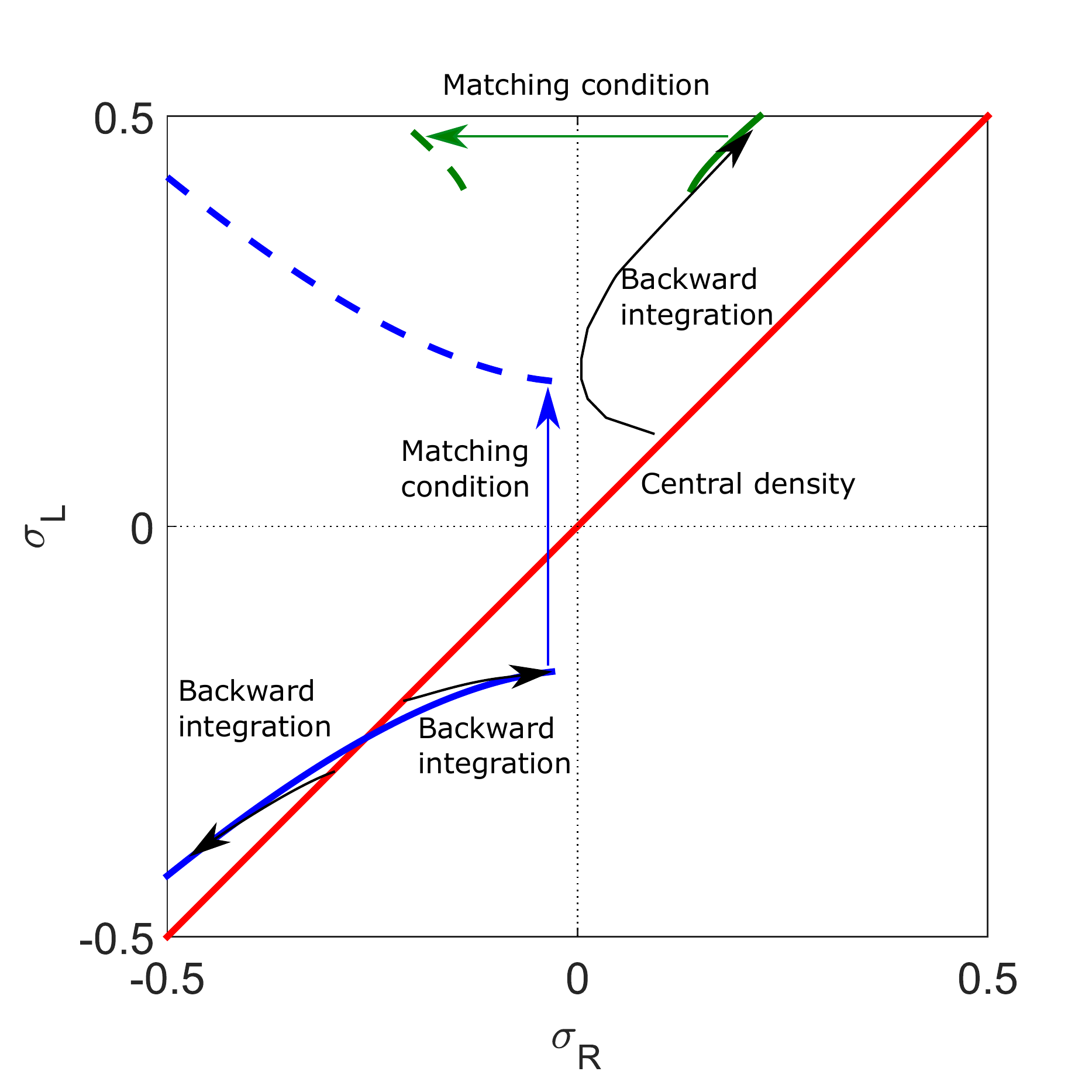}
\caption{Illustration of calculation of phase boundaries. The red line
  indicates the central density, which lies on the
  $\sigma_R = \sigma_L$ line for symmetric boundary
  conditions. Backward integration from this line (black arrows) to
  $x=-\frac{1}{2}$ gives the thick blue and green, where the blue line
  corresponds to $\sigma < 0$ and green line $\sigma > 0$. Applying
  the matching condition to $\sigma_L$ and $\sigma_R$ (blue and green
  arrows) gives the dashed blue and green lines, respectively. These
  are the phase boundaries of the LH phase. The green dashed line
  stops around $\sigma_R = -0.1386$ and won't extend to $\sigma_R = 0$
  line.  The switching rate is $0.5$\pers, bulk motor concentration
  $c = 200$ nM, and the motor speed $5$ \mum\pers; other parameters
  are the reference values of table \ref{tab:units}.}
\label{LHLH_phase_boundary_2}
\end{centering}
\end{figure}

\textbf{Boundaries of the LHLH phase:} When we integrate backwards
from the $\sigma_R=\sigma_L$ line, it is not possible to reach the
$\sigma_R = 0$ line if the switching rate is high
(fig. \ref{LHLH_phase_boundary}). The transition line then separates
the set of backwards integrated points into two regions (regions III
and IV described in sec.~\ref{sec:maximin},
fig. \ref{LHLH_phase_boundary_2}). The LHLH phase appears in the
region between $\sigma_R = 0$ and the LH phase boundary.

\begin{figure}[t!]
\begin{centering}
\includegraphics[width=0.45 \textwidth]{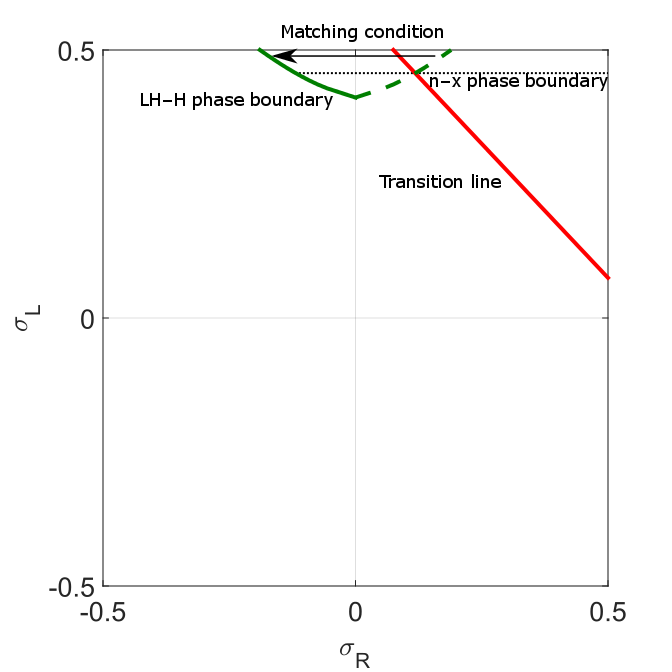}
\caption{Illustration of calculation of phase boundaries of local
  maximum and minimum phases.  The dashed green line is obtained from
  backward integration of the $\sigma_R = \sigma_L$ line to
  $x=-\frac{1}{2}$.  The green line indicates the LH phase boundary
  determined by the matching condition, where the right of the line is
  the H phase.  The red line is the transition line. Since the H phase
  satisfies the right boundary condition, the starting point of the
  density profile depends on $\sigma_L$ value. Therefore, the phase
  boundary of the local maximum and minimum phase is a horizontal line
  that intersects the green dashed line the and transition line. The
  local minimum phase is above the black dashed line. The switching
  rate is $0.1$\pers, bulk motor concentration $c = 200$ nM, and motor
  speed is $5$ \mum\pers; other parameters are the reference values of
  table \ref{tab:units}.}
\label{H_n_x_explanation}
\end{centering}
\end{figure}

\textbf{Boundaries of the H phase: } The region in the phase diagram
with high values of $1-\beta$ (small $\beta$) above the LH phase
corresponds to the high-density phase. In the H phase, the lane
left-end boundary condition is not satisifed. This phase is divided
into two regions in which the central density has a local maximun or
local minimum, controlled by the transition line. We determine the
separation between these two behaviors by determining where the set of
points $\sigma_{R,L} ( x=-\frac{1}{2})$ (integrated backwards from the
$\sigma_R=\sigma_L$ line) intercept the transition line
(fig. \ref{H_n_x_explanation}).

\subsection{Approximate phase boundaries using the total binding
  constraint}

\begin{figure}[t!]
\begin{centering}
\includegraphics[width=0.9 \textwidth]{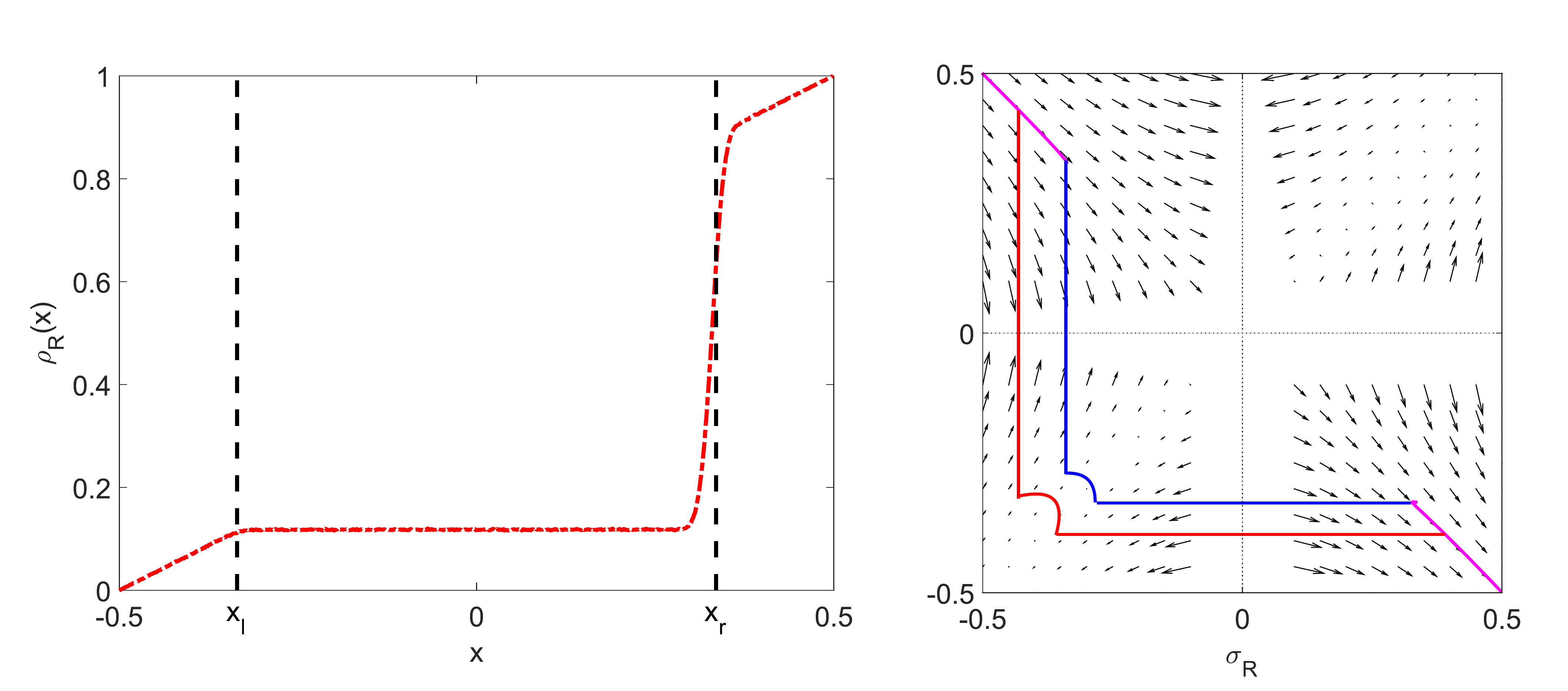}
\caption{Illustration of the density profile (left) and hypothetical
  phase-plane trajectories (right) in the LH phase.  Right: the purple
  curves indicate the density profile starting from the boundary
  conditions, and the blue and red curves are two different
  hypothetical trajectories which locally obey equations
  \eqref{eq:cont1}, \eqref{eq:cont2}, and \eqref{exact}. The
  parameters are $s = 0.44$\pers, $c = 200$ nM, $v = 5$ \mum\pers,
  $\kon = 2.7 \times 10^{-6}$ \pernm\pers, and $\koff = 0.00169$\pers;
  other parameters are the reference values of table \ref{tab:units}. Arrows in the right figure indicate vector field, which has the mathematical form in \eqref{flow_q+} and \eqref{flow_q-}.}
\label{one_case}
\end{centering}
\end{figure}

Phases that include domain walls mean that the density profile does
not always exactly satisfy the first-order mean-field steady-state
equations \eqref{eq:cont1}, \eqref{eq:cont2}, and \eqref{exact}. The
density is separated into several regions (fig. \ref{one_case}). In
principle, there could be multiple possible trajectories which satisfy
the boundary conditions and locally satisfy the differential equations
(fig. \ref{one_case} right). Despite the possibility of multiple
solutions for the same boundary conditions, our kMC simulation results
typically find just one stable steady-state solution for each set of
boundary conditions.

To understand this, we consider the number of equations and
unknowns. We focus on a single lane, e.g., the R lane. From equation
\eqref{real_space_q+}, we can derive the density profile starting from
the left end and the left boundary condition $\alpha_R$. We integrate
to the domain-wall position $x_l$; the corresponding density profile
is $\rho(x_l)$. We have two unknowns ($x_l$ and $\rho(x_l)$) and one
equation \eqref{real_space_q+}.  Using the same argument beginning
from the right end of the lane, we have two unknowns ($x_r$ and
$\rho(x_r)$) and one equation \eqref{real_space_q+}.  Then we can
determine the density profile between $x_l$ and $x_r$ using equation
\eqref{exact} or \eqref{real_space_q+} with boundary conditions
$\rho(x_l)$ and $1-\rho(x_r)$ \footnote{Note that the boundary
  condition $1-\rho(x_r)$ is due to the domain-wall matching
  condition.}. This adds no new variables and one equation
\eqref{real_space_q+}. Thus, there are four unknown variables: $x_r$,
$x_l$, $\rho(x_r)$, and $\rho(x_l)$, and three equations. We need one
more equation in order to uniquely determine the density profile. As
discussed above in
sec.~\ref{sec:Phase_space_flow_to_determine_the_density_profiles},
this can be done numerically using the finite-size
constraint. Alternatively, we can use the total binding constraint of
equation \eqref{eq:contin_constr}, as discussed in our previous work
\cite{kuan_motor_2015}.  Satisfying one of these constraints
automatically satisfies the other. The total binding constraint is
useful because it can be combined with analytic approximations to the
position-dependent density profile (equation \eqref{real_space_q+}).

Here we illustrate how to use the total binding constraint to derive
analytic estimates for the LH phase boundaries. For symmetric boundary
conditions, the domain-wall positions are also symmetric, so
$x_r=-x_{\rm bl}$ and $x_l=x_{\rm bl}$. If we consider the limit of
large motor speed, equations \eqref{flow_q+} and \eqref{flow_q-} can
be approximated by a piecewise linear form \cite{kuan_motor_2015}:
\begin{equation}
  \label{eq:rho_lin2}
  \rho_R(x) =  \begin{cases}  (\Konc+S)(x+\frac{1}{2})+\alpha  &
    -\frac{1}{2}\le x \le -x_{\rm bl}\\
    \frac{x}{2
      x_{\rm bl}}\left[(\frac{1}{2}-x_{\rm
        bl})(\Koff-\Konc)-\alpha+\beta\right]+ 
    \frac{1}{2}(\frac{1}{2}-x_{\rm bl})(\Konc+\Koff+2S)+\frac{\alpha+\beta}{2}&
    -x_{\rm bl} \le x \le x_{\rm bl}\\ 
     (\Koff + S)(x-\frac{1}{2})+1-\beta &  x_{\rm bl} \le x \le
     \frac{1}{2}  
  \end{cases}
\end{equation}
Using the total binding constraint of equation
\eqref{eq:contin_constr}, the integral of the density is
\begin{equation}
  \label{eq:domain_wall} 
  -\frac{1}{8} \left[4-\Koff+\Konc- 4 x_{\rm bl}^2 (3 \Koff + \Konc + 4 S) + 4
  \alpha - 4 \beta + 8 x_{\rm bl} (-1 + \Koff + S + 2 \beta)\right] =
  \rho_0+\frac{\alpha(1-\alpha)-\beta(1-\beta)}{\Konc + \Koff}. 
\end{equation}
The phase boundary between the LH and L phases occurs when the domain
wall position is $x_{\rm bl} = \frac{1}{2}$.  Then, equation
\eqref{eq:domain_wall} simplifies to
\begin{equation}
  \label{eq:alpha_beta} 
  \frac{\alpha+\beta}{2} =
  \rho_0+\frac{\alpha(1-\alpha)-\beta(1-\beta)}{\Konc + \Koff}, 
\end{equation}
or
\begin{equation}
  \label{eq:alpha_beta_2} 
  \beta = \frac{1}{4} \left[\Koff + \Konc + 2 \pm \sqrt{(\Koff + \Konc +
    2)^2 + 8 (\Koff + \Konc) \alpha - 16 \alpha(1-\alpha) - 16 (\Koff
    + \Konc) \rho_0}\right]. 
\end{equation}

Similarly, the phase boundary between the LH and H phases is found by
considering a piecewise-linear density profile like that above, but
with a jump to high density at $x_l$. In this case, the approximate
density profile is:
\begin{equation}
  \label{eq:rho_lin3}
  \rho_R(x) =  \begin{cases}  (\Konc+S)(x+\frac{1}{2})+\alpha  &
    -\frac{1}{2}\le x \le -x_{\rm bl}\\
    1-\frac{x}{2
      x_{\rm bl}}\left[(\frac{1}{2}-x_{\rm bl})(\Koff-\Konc)-\alpha+\beta\right] -
    \frac{1}{2}(\frac{1}{2}-x_{\rm bl})(\Konc+\Koff+2S)-\frac{\alpha+\beta}{2}&
    -x_{\rm bl} \le x \le x_{\rm bl}\\ 
    (\Koff + S)(x-\frac{1}{2})+1-\beta &  x_{\rm bl} \le x \le \frac{1}{2}  
  \end{cases}
\end{equation}
As above, the phase boundary between the LH and H phases occurs when
the domain wall position is $x_{\rm bl} = \frac{1}{2}$ (note that
$x_{\rm bl}$ is positive in our convention), giving a relation from
the total binding constraint of
\begin{equation}
  \label{eq:alpha_beta_3} 
  1-\frac{\alpha+\beta}{2} =
  \rho_0+\frac{\alpha(1-\alpha)-\beta(1-\beta)}{\Konc + \Koff}, 
\end{equation}
or

\begin{equation}
  \label{eq:alpha_beta_4}
  \beta = \frac{1}{4}\left[-\Koff - \Konc + 2 \pm \sqrt{(\Koff +
    \Konc-2)^2+16(\Koff + \Konc)-8 \alpha (\Koff +
    \Konc)-16\alpha(1-\alpha)-16\rho_0(\Koff + \Konc)}\right]
\end{equation}

\begin{figure}[t!]
\begin{centering}
  \includegraphics[width=0.45
  \textwidth]{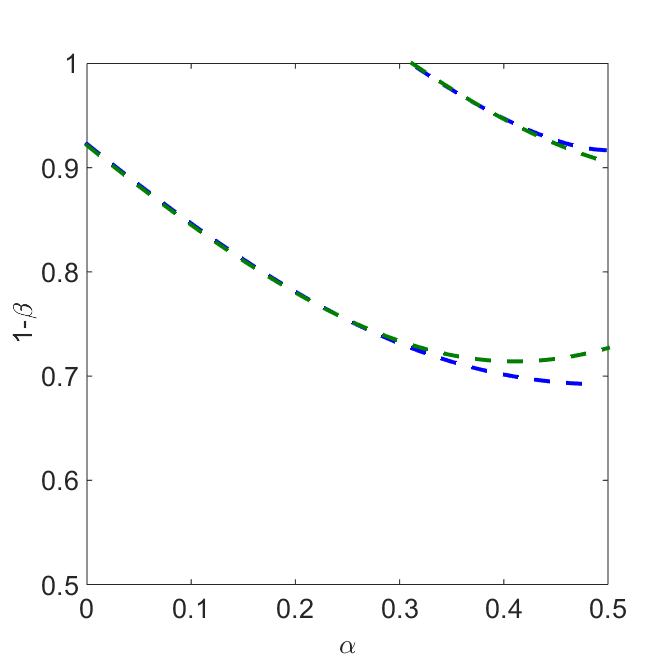}
  \caption{Comparison of determinination of the L boundaries using
    total binding constraint with linear approximation (green curves,
    equations \eqref{eq:rho_lin2} and \eqref{eq:rho_lin3}) and the
    finite-size constraint without any approximation (blue
    curves). The switching rate is $0.1$\pers, bulk motor
    concentration $c = 200$ nM, and motor speed $5$ \mum\pers; other
    parameters are the reference values of table \ref{tab:units}.}
\label{LH_boundary_tbc}
\end{centering}
\end{figure}

Figure \ref{LH_boundary_tbc} shows the result of determining the LH
phase boundaries using this approximation and the total binding
constraint. It agrees well with the numerically determined boundaries,
particularly for small $\alpha$.

\section{Nonequilibrium phases for general boundary conditions}
\label{sec:general_case}

\begin{figure}[t!]
\begin{centering}
\includegraphics[width=0.9 \textwidth]{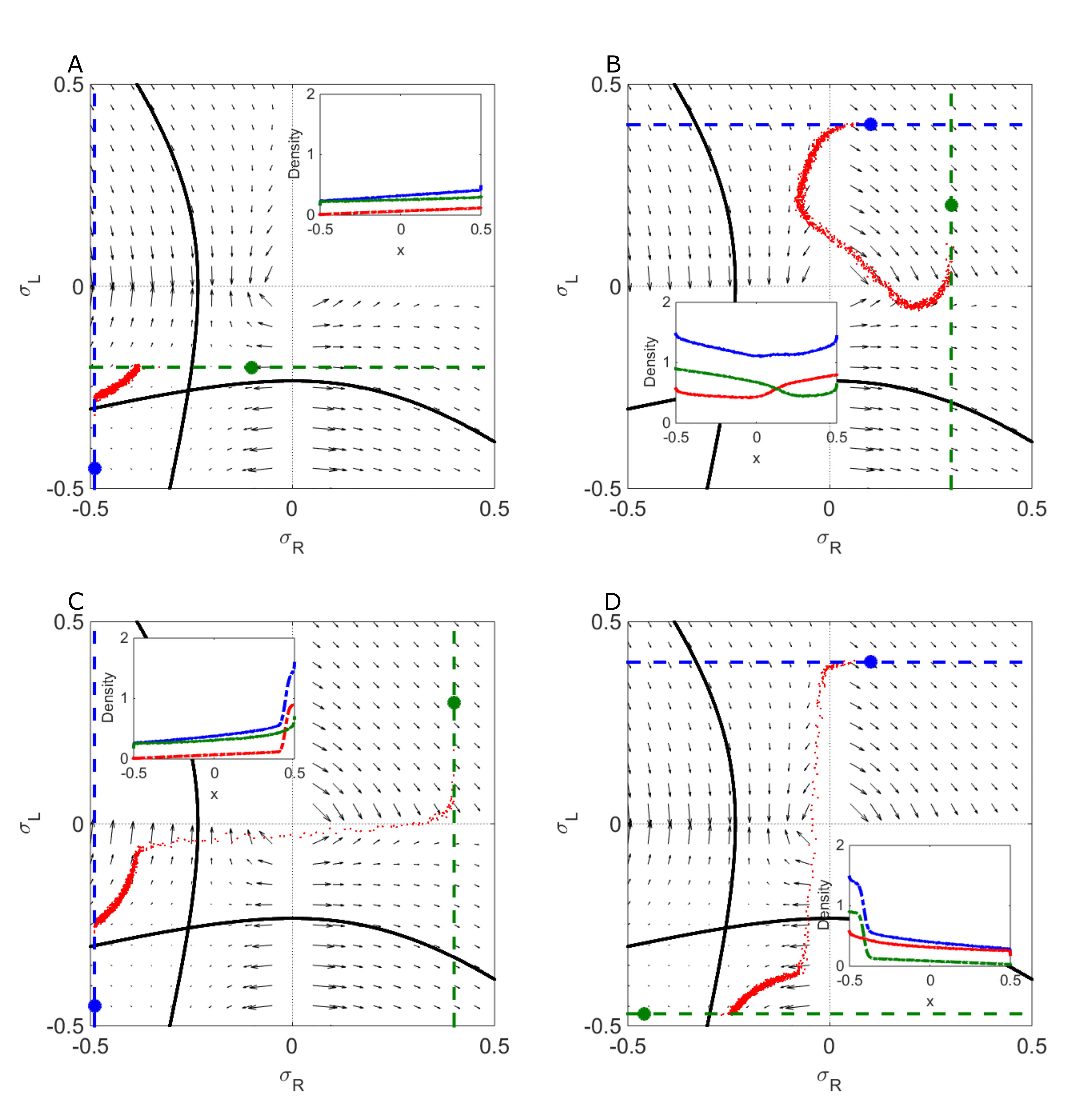}
\caption{Example phase-plane trajectories and density profiles (inset)
  for general boundary conditions. Insets show $\rho_R+\rho_L$ (blue),
  $\rho_R$ (red), and $\rho_L$ (green). Red dots are kMC simulation
  results. The blue dot indicates the left boundary condition and the
  green dot the right boundary condition.  A: The blue line is
  $\sigma_R = \alpha_R$, and the green line is $\sigma_L = \alpha_L$.
  B: The blue line is $\sigma_L = 1-\beta_R$, and the green line is
  $\sigma_R = 1-\beta_L$. C: The blue line is $\sigma_R = \alpha_R$,
  and the green line is $\sigma_R = 1-\beta_L$.  D: The blue line is
  $\sigma_L = 1-\beta_L$, and the green line is $\sigma_L = \alpha_L$.
  The switching rate is $0.1$\pers, bulk motor concentration $c = 200$
  nM, and motor speed $5$ \mum\pers; other parameters are the
  reference values of table \ref{tab:units}. Arrows indicate vector field, which has the mathematical form in \eqref{flow_q+} and \eqref{flow_q-}.}
\label{diff_bc}
\end{centering}
\end{figure}

\begin{figure}[t!]
\begin{centering}
\includegraphics[width=0.9 \textwidth]{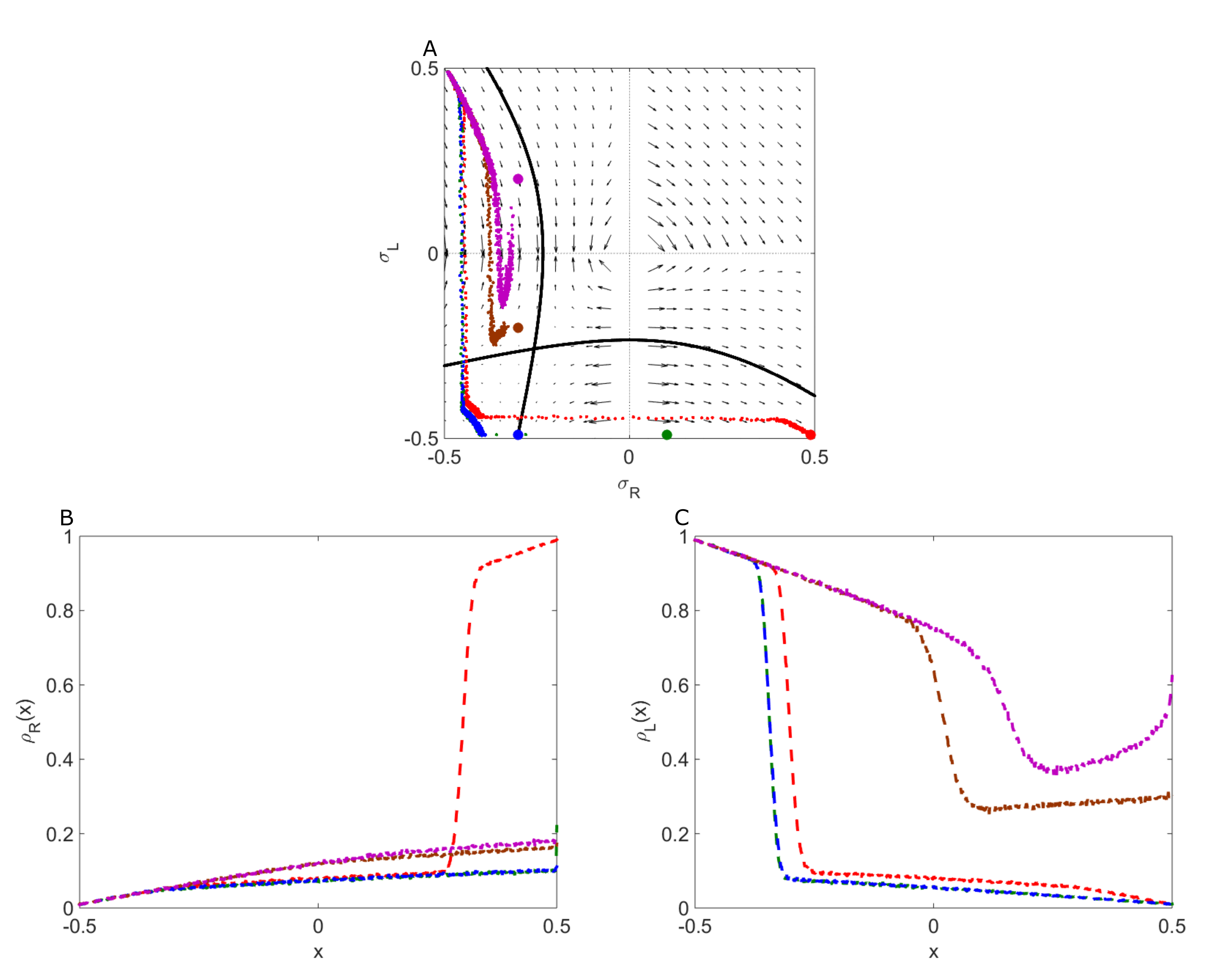}
\caption{Example phase-plane trajectories and density profiles for
  general boundary conditions, illustrating how altering the right
  boundary condition changes the density profile. A: Colored dots
  indicate the right boundary condition.  Red curve satisfies all four
  boundary conditions, green, blue, and brown curves do not satisfy
  $\beta_R$, and the purple curve satisfies $\alpha_R$ and
  $\beta_L$. B, C: corresponding density profiles on the R lane (B)
  and L lane (C). The switching rate is $0.1$\pers, bulk motor
  concentration $c = 200$ nM, and motor speed $5$ \mum\pers; other
  parameters are the reference values of table \ref{tab:units}. Arrows in figure A indicate vector field, which has the mathematical form in \eqref{flow_q+} and \eqref{flow_q-}.}
\label{move_ending}
\end{centering}
\end{figure}

In general, the boundary conditions might not symmetric for the two
lanes: $\alpha_R \neq \alpha_L$ and $\beta_R \neq \beta_L$.  In this
case, we cannot determine the density profile using the symmetry
argument that the central density lies on the $\sigma_R = \sigma_L$
line. However, we can still use the properties of the analytic
solution to the mean-field steady-state equations and the phase space
flow to determine properties of the solutions.  The density profile
locally follows the phase space flow, and is connected by a curve with
the correct number of sites that links the left boundary condition
$\sigma_R(x=-\frac{1}{2}) =\alpha_R-\frac{1}{2}$,
$\sigma_L(x=-\frac{1}{2})=\frac{1}{2}-\beta_L$ to the right boundary
condition $\sigma_R(x=\frac{1}{2}) =\frac{1}{2}-\beta_R$,
$\sigma_L(x=\frac{1}{2})=\alpha_L-\frac{1}{2}$. The matching condition
for the domain wall can be applied, if necessary. We can describe the
possible behaviors based on whether or not each of the four boundary
conditions if satisfied. This gives $2^4=16$ possible cases, which can
be grouped into ten classes illustrated in figs.~\ref{diff_bc} and
\ref{move_ending}.

The location of the boundary conditions in the phase plane determines
the types of behavior that can occur. We will therefore consider which
quadrant in the phase plane (upper right, upper left, lower left,
lower right) contains the left
$(\alpha_R-\frac{1}{2}, \frac{1}{2}-\beta_L)$ and right
$(\frac{1}{2}-\beta_R, \alpha_L-\frac{1}{2})$ boundary conditions.

\textbf{All boundary conditions satisfied:} This case is analogous to
the LH phase in the symmetric phase diagram. It often occurs when the
left boundary condition $(\alpha_R-\frac{1}{2}, \frac{1}{2}-\beta_L)$
is in the upper left quadrant of the phase plane and the right
boundary condition $(\frac{1}{2}-\beta_R, \alpha_L-\frac{1}{2})$ is in
the lower right quadrant. A domain wall occurs either in the upper
right or lower left quadrant.

\textbf{No boundary conditions satisfied:} This case is analogous to
the M phase in the symmetric phase diagram. It often occurs when the
left boundary condition is in the lower right quadrant and the right
boundary condition is in the upper left quadrant. The density profile
is independent of the boundary conditions.

\textbf{Both lane minus-end boundary conditions satisfied:} Thus case
is analogous to the L phase in the symmetric phase diagram. It often
occurs when the left boundary condition and the right boundary
condition are in the lower left quadrant. The density profile follows
a trajectory which obeys
$\sigma_R(x=-\frac{1}{2}) = \alpha_R-\frac{1}{2}$,
$\sigma_L(x=\frac{1}{2}) = \alpha_L-\frac{1}{2}$, and contains the
correct number of sites (fig. \ref{diff_bc}A).

\textbf{Both lane plus-end boundary conditions satisfied:} This case
is analogous to the H phase in the symmetric phase diagram. It often
occurs when the left boundary condition and the right boundary
condition are in the upper right quadrant. The density profile follows
a trajectory which obeys
$\sigma_R(x=\frac{1}{2}) = \frac{1}{2}-\beta_R$,
$\sigma_L(x=-\frac{1}{2}) = \frac{1}{2}-\beta_L$, and contains the
correct number of sites( fig. \ref{diff_bc}B).

\textbf{Both lane minus-end and one plus-end boundary conditions
  satisfied:} This case is analogous to a semi-LH phase. It can occur
two ways, depending on whether lane L or R has its plus-end boundary
condition satisfied.  The L case often occurs when the left boundary
condition is in the upper left quadrant and the right boundary
condition is in the lower left quadrant. This phase occurs when the
right boundary point moves from the lower right quadrant (where all
boundary conditions can be satisfied) to the lower left quadrant.  The
domain wall of the R lane moves beyond $x_w=\frac{1}{2}$, so that the
right boundary condition for the R lane is not satisfied.

The R case is symmetric with the L case. It occurs when the left
boundary condition is in the lower left quadrant and the right
boundary condition is in the lower right quadrant. This phase occurs
when the left boundary point moves from the the upper left quadrant
(where all boundary conditions can be satisfied) to the lower left
quadrant. The domain wall of the L lane moves beyond
$x_w=-\frac{1}{2}$, so that the left boundary condition for the L lane
is not satisfied.

\textbf{Left or right boundary conditions satisfied:} This case is
analogous to a semi-LH phase. First we consider when the left boundary
condition is satisfied. This often occurs when both the left and right
boundary conditions are in the upper left quadrant.  This case can be
treated as moving the right boundary condition from the lower left
quadrant to the upper left quadrant. Once the right boundary condition
passes the $\sigma_L = 0$ line, the flow cannot reach the right
boundary condition, even with a domain wall. Therefore, the density
profile is dominated by the left boundary condition only.

The case in which the right boundary condition is satisfied is
symmetric. This often occurs when both boundary conditions are in the
lower right quadrant. This case can be treated as moving the left
boundary condition from the lower left quadrant to the lower right
quadrant. Once the right boundary condition passes the $\sigma_R = 0$
line, the flow cannot reach the left boundary condition, even with a
domain wall. Therefore, the density profile is dominated by the right
boundary condition only.

\textbf{Only one minus-end boundary condition satisfied:} This case is
analogous to a semi-LH phase.  The case in which the right lane
minus-end boundary condition $\alpha_R$ is satisfied often occurs when
the left boundary condition is in the lower left quadrant and the
right boundary condition is located in the upper left quadrant. This
case can be treated as moving the left boundary condition from the
upper left quadrant to the lower left quadrant. Once the domain wall
position in the left lane is less than $-\frac{1}{2}$, the left
boundary condition of the left lane is not satisfied. Therefore, the
density profile is dominated by the R lane minus-end boundary
condition only.

The case in which the left-lane minus-end boundary condition
$\alpha_L$ is satisfied is symmetric. This often occurs when the left
boundary condition is in the lower right quadrant and the right
boundary condition is in the lower left quadrant. This case can be
treated as moving the right boundary condition from the lower right
quadrant to the lower left quadrant. Once the domain wall position in
the right lane is greater than $\frac{1}{2}$, the right boundary
condition of the right lane is not satisfied. Therefore, the density
profile is dominated by the L lane minus-end boundary condition only.

\textbf{Both lane plus-end and one minus-end boundary conditions
  satisfied:} This case is analogous to a semi-LH phase. The case in
which all but $\alpha_L$ is satisfied often occurs when the left
boundary condition is in the upper left quadrant and the right
boundary condition is in the upper right quadrant. This case can be
treated as moving the right boundary condition from the lower right
quadrant to the upper right quadrant. Once the domain wall position in
the left lane is greater than $\frac{1}{2}$, the right boundary
condition of the left lane is not satisfied.

The case in which all but $\alpha_R$ is satisfied is symmetric. This
case often occurs when the left boundary condition is in the upper
right quadrant and the right boundary is in the lower right
quadrant. This case can be treated as moving the left boundary
condition from the upper left quadrant to the upper right
quadrant. Once the domain wall position in the right lane is less than
$-\frac{1}{2}$, the left boundary condition of the right lane is not
satisfied.

\textbf{Both boundary conditions on one lane satisfied:} The case in
which the R lane boundary conditions are satisfied often occurs when
the left boundary condition is in the lower left quadrant and the
right boundary condition is in the upper right quadrant. The density
profile follows a trajectory which obeys
$\sigma_R(x=-\frac{1}{2}) = \alpha_R-\frac{1}{2}$,
$\sigma_R(x=\frac{1}{2}) = \frac{1}{2}-\beta_R$, and contains the
correct number of sites (fig.~\ref{diff_bc}C).

The case in which the L lane boundary conditions are satisfied is
symmetric.  This case often occurs when the left boundary condition is
in the upper right quadrant and the right boundary condition is in the
lower left quadrant. The density profile follows a trajectory which
obeys $\sigma_L(x=-\frac{1}{2}) = \frac{1}{2}-\beta_L$,
$\sigma_L(x=\frac{1}{2}) = \alpha_L-\frac{1}{2}$, and contains the
corred number of sites (fig. \ref{diff_bc}D).

\textbf{Only one plus-end boundary condition satisfied:} First we
consider when the R lane plus-end boundary condition is satisfied.
This case often occurs when the left boundary condition is in the
upper right quadrant and the right boundary condition is in the upper
left quadrant. This case can be treated as moving the right boundary
condition from the upper right quadrant to the upper left
quadrant. Once the right boundary condition crosses $\sigma_R = 0$,
the L lane plus-end boundary condition is not satified.

The case when the L lane plus-end boundary condition is satisfied is
symmetric. This case often occurs when the left boundary condition is
in the lower right quadrant and the right boundary condition is in the
upper right quadrant. This case can be treated as moving the left
boundary condition from the upper right quadrant to the lower right
quadrant. Once the right boundary condition crosses $\sigma_L = 0$,
the R lane plus-end boundary condition cannot be satisfied.

\section{Conclusion}
\label{sec:conclusion}

We have studied a model of the TASEP on two antiparallel lanes with
Langmuir kinetics and lane switching (fig.~\ref{cartoon}). We define
the model and derive the mean-field continuum equations and the total
binding constraint, as well as the kinetic Monte Carlo simulation
rules, in sec.~\ref{sec:model}. In table \ref{tab:units}, we list the
reference parameter set measured or estimated from the BTS experiments
\cite{bieling_minimal_2010} and our previous work
\cite{kuan_motor_2015}.

Since the steady-state mean-field equations are nonlinear and strongly
coupled (for sufficiently high switching rate), we study their
solutions in the density-density phase plane (sec.~\ref{sec:phase}).
We find an analytical solution in the phase plane and an expansion to
determine position-dependent approximate solutions. Studying the phase
space flow and fixed points of the model
(fig.~\ref{nonlinear_phase_space_flow}) gives intuition for the phases
and how they change with parameters. In particular, both the number
and location of the phase-plane fixed points (fig.~\ref{fixed_points})
change with switching rate: for sufficiently high switching rate, two
additional fixed points appear, leading to qualitative changes in the
behavior of the model. This allows a new multi-phase coexistence low
density-high density-low density-high density (LHLH) phase to appear.
In the mean-field model, we can calculate exactly the critical
switching rates at which these changes occur. In addition, phase plane
analysis allows us to determine domain wall positions using the
finite-size constraint (fig.~\ref{example_flow}).

We then use the phase-plane analysis to determine the nonlinear phases
that can occur for the case of symmetric boundary conditions
(sec.~\ref{sec:sym_phase_diagram}). The fixed points divide the lanes'
central density into 4 regions (fig.~\ref{Regions}) with different
flow properties. These determine the phases that can occur. For low
switching rate, the low-density (L), high-density (H), low
density-high density coexistence (LH), and Meissner (M) phases
previously studied by PFF for the single-lane case
\cite{parmeggiani_totally_2004} occur
(fig.~\ref{nonlinear_phase_space_flow_3}). For high switching rate,
the LHLH phase appears (fig.~\ref{nonlinear_phase_space_flow_4}). We
also determine which boundary conditions are satisfied in different
phases (fig.~\ref{LD_HD_boundary_stability}).

The analysis of the phase-space flows and fixed points allow us to
determine the phase diagram for the biophysically relevant case of
symmetric boundary conditions (sec.~\ref{sec:phase_diagram}).
Fig.~\ref{phase_diagram} illustrates the phase diagrams for low and
high switching rate. We then discuss the calculation of the boundaries
of each phase, particularly using backward integration from the lanes'
center to determine the LH phase boundaries
(fig.~\ref{LH_phase_boundary}), its changes with motor speed
(fig.~\ref{LH_phase_boundary_hl}), and the LHLH phase boundaries
(figs.~\ref{LHLH_phase_boundary}, \ref{LHLH_phase_boundary_2}). A
similar method can be used to determine whether there is a local
maximum or minimum at the lanes' center
(fig.~\ref{H_n_x_explanation}). Additionally, we discuss an alternate
method for determining approximate phase boundaries of the LH phase
using an analytic approximation to the density profile
(fig.~\ref{one_case}) and the total binding constraint. The
approximate phase boundaries computed in this way are close to those
determined from the phase-plane analysis (fig.~\ref{LH_boundary_tbc}).

Finally, we considered the general case of asymmetric boundary
conditions (sec.~\ref{sec:general_case}). There are 10 cases
corresponding to different possibilities for which boundary conditions
are satisfied; we show some examples in fig.~\ref{diff_bc}.  Changes
in the location of the boundary point in the phase plane cause
predictable changes (fig.~\ref{move_ending}).

For our model of a TASEP with two antiparallel lanes and binding and
switching kinetics, the phase-plane analysis we describe is useful
because the analytic solution to the mean-field steady-state equations
allows us to determine the trajectories and fixed points. This
approach may be useful in the future for the study of other multi-lane
TASEP models. Because motors can regulate microtubule length and other
biochemical reactions \cite{varga_kinesin8_2009, bieling_minimal_2010,
  goshima_length_2005, walczak_xkcm1_1996}, this method can be used to
predict how experimental parameters might alter biochemical activity
through alterations in motor density distributions. In particular, our
ability to predict the spatial distribution of motor accumulation
(particularly in the LH and LHLH phase) in an antiparallel overlap
might be important for antiparallel overlap length regulation during
mitosis \cite{bieling_minimal_2010}. In the future, it might be of
interest to consider how the model we consider would change if the two
lanes have different motor properties (in binding kinetics, motor speed, or
switching rate), or if motor properties change spatially along a
lane. Either of these two cases could occur due to tubulin
post-translational modifications, which can differentially alter motor
interactions with MTs carrying modifications  \cite{janke_post_2011}.

\begin{acknowledgements}
  We thank Robert Blackwell, Matthew Glaser, and Loren Hough for
  useful discussions. This work was supported by NSF grants
  DMR-0847685 and DMR-1551095 and NIH grant K25GM110486 to MDB,
  fellowship to H-SK provided by matching funds from the NIH/CU
  Biophysics Training Program, and facilities of the Soft Materials
  Research Center under NSF MRSEC grant DMR-1420736.
\end{acknowledgements}

%\bibliography{references}{}


\end{document}